\documentclass[%
 reprint,
 amsmath,amssymb,
 aps,
]{revtex4-2}
\usepackage{tikz}
\usepackage{amsmath, amssymb}
\usepackage{simpler-wick}
\usepackage{graphicx}
\usepackage{dcolumn}
\usepackage{bm}

\usepackage{hyperref} 
\usepackage{lipsum}
\usepackage[T1]{fontenc} 
\usepackage{slashed}
\usepackage{arydshln}
\usepackage{caption}
\usepackage{subcaption}
\usepackage{nicematrix}
\usepackage{tkz-euclide}
\usepackage[compat=1.1.0]{tikz-feynman}
\usepackage{booktabs}
\usepackage{setspace}
\usepackage{hhline}
\usepackage{stackengine}
\usepackage{multirow}
\usepackage{array}
\usepackage{adjustbox}
\usepackage{makecell}
\usepackage{pifont}
\usepackage{nccmath}

\begin{document}

\preprint{APS/123-QED}

\title{Probing Excited $q\bar{q}$ Mesons via QCD Sum Rules}

\author{Shuang-Hong Li}
\email{shlee@zju.edu.cn}
\affiliation{Zhejiang Institute of Modern Physics, School of Physics, Zhejiang University, Hangzhou, 310027, China}

\author{Wei-Yang Lai}
\email{laiweiyang@zju.edu.cn}
\affiliation{Zhejiang Institute of Modern Physics, School of Physics, Zhejiang University, Hangzhou, 310027, China}

\author{Hong-Ying Jin}
\email{jinhongying@zju.edu.cn}
\affiliation{Zhejiang Institute of Modern Physics, School of Physics, Zhejiang University, Hangzhou, 310027, China}

\date{\today}

\begin{abstract}

We present a systematic study of the masses of light excited $q\bar{q}$ mesons using QCD sum rules at next-to-leading order (NLO). To probe excited states, we construct several interpolating currents with covariant derivatives inserted. The calculation is carried out up to dimension-8 condensates, including NLO perturbative and $m\langle\bar{q}q\rangle$ corrections. Employing Gaussian sum rules, we obtain several $J^P=2^\pm$ nonets with masses that agree well with experiments. Several $J=0,1$ states compatible with experiments are also obtained using both Gaussian and Laplace sum rules. In particular, the $J^{PC}=2^{++}$ current couples to two distinct $2^{++}$ resonances. This work demonstrates the efficacy of operators with covariant derivatives for studying excited hadrons.

\end{abstract}

\maketitle


\section{Introduction\label{sec_intro}}

The $q\bar{q}$ mesons are the simplest hadrons; understanding their spectrum is a central problem in hadron physics. Although the properties of ground-state $q\bar{q}$ mesons are described reasonably well by quark models, the excited spectrum still poses open questions regarding assignments, decay constants, and mixing. Furthermore, there is growing evidence that some excited $q\bar{q}$ mesons mix with exotic states (non-$q\bar{q}$ mesons)~\cite{mixing_x_y,X3872,chen_xyz,X3872_molecule-charmonium}, making the excited spectrum more intricate and challenging to interpret.

On the other hand, QCD sum rules~\cite{qsr,qsr_book,qsr_intro, qsr_intro_2,qsr_laplace}, which connect hadron properties to the QCD Lagrangian and the nonperturbative QCD vacuum (encoded in the so-called condensates), usually focus on the ground state; only sporadic investigations have been devoted to excited states. This is because the ground state is naturally favored in QCD sum rule analyses: the interpolating operator for the ground state can be directly constructed, and the lowest-lying state in the spectrum will dominate after Borel transformation~\cite{qsr,qsr_intro,qsr_book}. To investigate radially or orbitally excited $q\bar{q}$ mesons, it is necessary to design operators that couple more strongly to excited states.

One approach to constructing such operators is by inserting covariant derivatives into the $q\bar{q}$ meson operator; the $\overline{\Psi}\Gamma{\small\overleftrightarrow{\nabla}^n}\Psi$-type operator is expected to couple preferentially to excited $q\bar{q}$ mesons due to the extra momentum weight. Here, $\Gamma$ denotes a generic Dirac gamma matrix and ${\small\overleftrightarrow{\nabla}=\nabla-\overleftarrow{\nabla}}$, where $\nabla$ is the covariant derivative in the fundamental representation. In momentum space, ${\small\overleftrightarrow{\nabla}}$ corresponds to the quarks' relative momentum; thus, the coupling between $\overline{\Psi}\Gamma{\small\overleftrightarrow{\nabla}^n}\Psi$ and the S-wave $q\bar{q}$ meson should be suppressed. Operators with explicit covariant derivatives have already been used in several works~\cite{,REINDERS1982125, ALIEV1982401,REINDERS19851, Kim1997, Kataev_2005, PhysRevD.79.074025,HadronSpectrum_2012gic, Du_2013, PhysRevD.106.014023,J_3-,lsr_qdq}, and have yielded some successful results.

In this work, we present an NLO QCD sum rule investigation of light excited $q\bar{q}$ meson masses using the $\overline{\Psi}\Gamma{\small\overleftrightarrow{\nabla}}\Psi$ operator. This paper is organized as follows: the construction of currents is described in Section~\ref{currents_basis}; the Gaussian sum rule analysis is presented in Section~\ref{num} and concluded in Section~\ref{conclusion}; the Laplace sum rule results are supplemented in Appendix~\ref{sec_lsr}; the evaluation of dimension-8 condensates is presented in Appendix~\ref{d7_8_condensate}.

\section{Operator Basis, Quantum Numbers, and Projectors\label{currents_basis}}

To calculate the correlator of interpolating currents in the low-energy region, the operator product expansion (OPE) is utilized in QCD sum rules~\cite{qsr,qsr_book,qsr_intro,qsr_intro_2,qsr_laplace}:
\begin{equation}
	\begin{split}
	\Pi_{AB}(q^2)&=i\int d^4x\, e^{-iqx}\langle T J_A(x)J_B^\dagger(0)\rangle\\
	&= \sum_i \mathcal{T}^i_{AB}\,\Pi_i(q^2),
	\end{split}
\end{equation}
where $J_A$ and $J_B$ are specific currents coupled to the state under investigation; $\mathcal{T}^i_{AB}$ is the Lorentz (tensor) projector corresponding to a specific spin, and $\Pi_i(q^2)$ is a scalar function of $q^2$; contributions with different spins are involved in general. The correlator and spectral function are related through the dispersion relation~\cite{qsr,qsr_book,qsr_intro, qsr_intro_2}:
\begin{equation}
	\frac{1}{\pi}\text{Im}\Pi_i(s) \sim f_i^2\delta(s-m_i^2)+\theta(s-s_0)\rho_i(s).
	\label{pole_continue}
\end{equation}
Here, the ``narrow resonance $+$ continuum'' ansatz~\cite{qsr,qsr_book,qsr_intro, qsr_intro_2} for the spectrum is adopted; $m_i$ is the mass of the lowest-lying resonance; $s_0$ is the continuum threshold; and $\rho_i(s)$ is the continuum spectral density. For completeness, we employ five $\overline{\Psi}\Gamma{\small\overleftrightarrow{\nabla}}\Psi$-type operators:
\begin{equation}
	\begin{split}
		J^\mu &= \overline{\Psi}_{f_a}\overleftrightarrow{\nabla}^\mu\Psi_{f_b},\\
		\widetilde{J}^\mu &= \overline{\Psi}_{f_a}\gamma^5\overleftrightarrow{\nabla}^\mu\Psi_{f_b},\\
		J^{\mu\alpha} &= \overline{\Psi}_{f_a}\gamma^\mu \overleftrightarrow{\nabla}^\alpha\Psi_{f_b},\\
		\widetilde{J}^{\mu\alpha} &= \overline{\Psi}_{f_a}\gamma^5\gamma^\mu\overleftrightarrow{\nabla}^\alpha\Psi_{f_b},\\
		J^{\mu\nu\alpha} &= \overline{\Psi}_{f_a}\sigma^{\mu\nu}\overleftrightarrow{\nabla}^\alpha\Psi_{f_b},
		\label{current_basis}
	\end{split}
\end{equation}
where $f_a$ and $f_b$ are flavor labels and $\sigma^{\mu\nu}=\frac{i}{2}[\gamma^\mu,\gamma^\nu]$.

The currents $\widetilde{J}^\mu$, $J^{\mu\alpha}$, and $\widetilde{J}^{\mu\alpha}$ in Eq.~\eqref{current_basis} were already considered in the early QCD sum-rule literature~\cite{REINDERS1982125, ALIEV1982401, REINDERS19851}, albeit in different notations. Those early studies established that derivative currents can interpolate non-$S$-wave mesons, but they treated only a few selected channels, focusing on the highest allowed spin for each current. The present work extends that program by organizing these currents into a unified basis and deriving the full set of Lorentz projectors and accessible quantum numbers. Additionally, a more systematic OPE and sum-rule analysis at NLO is performed, and the Gaussian sum rule is adopted for detailed numerical analysis.

Some configurations in Eq.~\eqref{current_basis} yield conserved currents. The current $\frac{1}{2}\,(\overline{\Psi}\gamma^\mu \overleftrightarrow{\nabla}^\alpha\Psi - \overline{\Psi}\gamma^\alpha \overleftrightarrow{\nabla}^\mu\Psi)\equiv\overline{\Psi}\gamma^{[\mu} \overleftrightarrow{\nabla}^{\alpha]}\Psi$ is conserved as a consequence of the conservation of the canonical energy-momentum tensor~\cite{refId0,PhysRevD.106.125012}:
\begin{equation}
	\partial_\mu \Big(\frac{i}{2}\overline{\Psi}\gamma^\mu \overleftrightarrow{\nabla}^\alpha\Psi-2\text{Tr}[{F^\mu}_\beta F^{\alpha\beta}]-g^{\mu\alpha}\mathcal{L}\Big)=0.
\end{equation}
In the massless limit, the left- and right-handed spinors are decoupled, and $\overline{\Psi}\gamma^5\gamma^{[\mu} \overleftrightarrow{\nabla}^{\alpha]}\Psi$ is conserved as well. The conservation of $\overline{\Psi}\overleftrightarrow{\nabla}^\mu\Psi$ is easy to verify using the Gordon decomposition~\cite{Gordon}:
\begin{equation}
	2m\overline{\Psi}\gamma^\mu\Psi=i \big(\overline{\Psi}\overleftrightarrow{\nabla}^\mu\Psi +\partial_\nu \overline{\Psi}\sigma^{\mu\nu}\Psi\big),
\end{equation}
where $\overline{\Psi}\gamma^\mu\Psi$ is conserved due to global $U(1)$ symmetry. A straightforward calculation shows that $\overline{\Psi}\gamma^5\overleftrightarrow{\nabla}^\mu\Psi$ is also conserved, as long as $\gamma^5$ and $\gamma^\mu$ are treated as anti-commuting:
\begin{equation}
	\begin{split}
	\partial_\mu \big(\overline{\Psi}\gamma^5\overleftrightarrow{\nabla}^\mu\Psi\big)=&\,\,\overline{\Psi}\overleftarrow{\partial}_\mu\gamma^5\nabla^\mu\Psi + \overline{\Psi}\gamma^5\partial_\mu\nabla^\mu\Psi\\
	&-\overline{\Psi}\overleftarrow{\nabla}^\mu\overleftarrow{\partial}_\mu\gamma^5\Psi - \overline{\Psi}\overleftarrow{\nabla}^\mu\gamma^5\partial_\mu\Psi\\
	=&\,\, g^{\mu\nu}\big(\overline{\Psi}\gamma^5\nabla_\mu\nabla_\nu\Psi \!-\! \overline{\Psi}\overleftarrow{\nabla}_\nu\overleftarrow{\nabla}_\mu\gamma^5\Psi\big)\\
	=&\,\,\frac{1}{2}\big(\overline{\Psi}\gamma^5\sigma^{\mu\nu}G_{\mu\nu}\Psi \!-\! \overline{\Psi}G_{\mu\nu}\sigma^{\mu\nu}\gamma^5\Psi\big)\\
	=&\,\,0
	\end{split}
	\label{J5_conserve}
\end{equation}
Here, the quark equation of motion is used. The conservation of $\overline{\Psi}\gamma^{[\mu} \overleftrightarrow{\nabla}^{\alpha]}\Psi$ can also be derived following a calculation similar to Eq.~\eqref{J5_conserve}.

Since the fundamental QCD coupling is flavor-blind, the differences between quarks are manifested in their masses. Similar conservation for $\overline{\Psi}_{f_a}\Gamma{\small\overleftrightarrow{\nabla}}\Psi_{f_b}$ with $f_a\neq f_b$ should also hold if $m_{f_a}=m_{f_b}$. Up to $O(\alpha_s)$ in this work, $J^\mu$, $\widetilde{J}^\mu$, and $J^{[\mu\alpha]}$ are conserved when $m_{f_a}=m_{f_b}$, while $\widetilde{J}^{[\mu\alpha]}$ is conserved when $m_{f_a}=m_{f_b}=0$. In the massless limit, the correlators of these currents have no longitudinal perturbative contributions. At $O(\alpha_s^2)$, the diagram\vspace{-0.25cm}
\begin{equation*}
	\begin{tikzpicture}[baseline=-\the\dimexpr\fontdimen22\textfont2\relax]
		\begin{feynman}
			\vertex (al);
			\vertex [right=2.4cm of al](ar);
			\vertex [right=0.9cm of al](ml);
			\vertex [right=1.5cm of al](mr);
			\vertex [above=0.3cm of ml](mlu);
			\vertex [above=0.3cm of mr](mru);
			\diagram*[small]{
				(ar)-- [bend right=25] (mru)--[bend right=25](ar),
				(al)-- [bend right=25] (mlu)--[bend right=25](al),
				(mlu)--[photon, bend left = 10](mru),
				(al)--[photon, bend right = 36](ar),
			};
		\end{feynman}
	\end{tikzpicture}
\end{equation*}
is involved only for $f_a = f_b$, and is absent for $f_a\neq f_b$. Due to this difference, the conservation of $J^\mu$, $\widetilde{J}^\mu$, $J^{[\mu\alpha]}$ and $\widetilde{J}^{[\mu\alpha]}$ beyond $O(\alpha_s)$ is not expected when $f_a\neq f_b$.

The quantum numbers of $q\bar{q}$ mesons follow a specific pattern. For a $q\bar{q}$ state with orbital angular momentum $l$ and spin $s$, the parity is $P=(-1)^{(l+1)}$, while the C-parity is $C=(-1)^{(l+s)}$. Therefore, $q\bar{q}$ states with quantum numbers $J^{PC}=0^{++}, 1^{+\pm}$, and $J\geq2$ are necessarily excited states. To isolate information about specific channels from the correlator, one must write the complete basis of polarization tensors $\epsilon(p)$ in
\begin{equation}
	\langle 0|J|h(p)\rangle=\epsilon(p)f(p^2)
\end{equation}
for each current $J$ and hadron state $|h(p)\rangle$, where $f(p^2)$ is the coupling strength. The polarization tensors for each current in Eq.~\eqref{current_basis} are listed in Table~\ref{polar_summary}. In the rest frame, each polarization tensor corresponds to an irreducible representation space of the rotation group. To obtain the corresponding Lorentz projectors, the following relations are used~\cite{Tanju_spin-2,Jafarzade_spin-3}:
\begin{equation}
	\begin{split}
		\sum_\lambda\epsilon_\lambda^\mu \epsilon_\lambda^{*\nu}&=\frac{p^\mu p^\nu}{p^2}-g^{\mu\nu}\equiv\eta^{\mu\nu},\\
		\sum_\lambda\epsilon_\lambda^{\mu\alpha}\epsilon_\lambda^{*\nu\beta}\!&=-\frac{1}{d-1}\eta^{\mu\alpha}\!\eta^{\nu\beta}\!\!+\!\frac{1}{2}\Big(\eta^{\mu\nu}\!\eta^{\alpha\beta}\!+\!\eta^{\mu\beta}\!\eta^{\nu\alpha}\Big),
		\label{polarization-2-3}
	\end{split}
\end{equation}
where $\epsilon_\lambda^\mu$ and $\epsilon_\lambda^{\mu\alpha}$ are spin-1 and spin-2 polarization tensors, respectively; $d$ is the dimension of spacetime. The relation for the Levi-Civita tensor
\begin{equation}
	\epsilon^{\mu\nu\alpha\beta}\epsilon_{\mu^\prime\nu^\prime\alpha^\prime\beta^\prime}=4!\,{g^\mu}_{[\mu^\prime}\,{g^\nu}_{\nu^\prime}\,{g^\alpha}_{\alpha^\prime}\,{g^\beta}_{\beta^\prime]}
	\label{levi_civita_sum}
\end{equation}
is also used.

\begin{table}[t!]
	\centering
	\caption{Summary of the LO perturbative and $\langle GG\rangle$ contributions for each configuration. The second column lists the allowed quantum numbers with polarization tensors, where square and curly brackets denote anti-symmetrization and symmetrization of the indices, respectively; the corresponding Lorentz projectors can be obtained using Eqs.~\eqref{polarization-2-3} and \eqref{levi_civita_sum}. The ``$\times$'' means no such contribution; ``0'' means the contribution exists but vanishes in the massless limit; ``$-$'' means the contribution exists but does not contribute to the imaginary part of the correlator (for dimensionless Lorentz projectors).\label{polar_summary}}
	\renewcommand{\arraystretch}{1}
	\begin{ruledtabular}
		\begin{tabular}{clcc}
			&&LO perturbative $\,$ &$\langle GG\rangle$\\
			\hline
			\multirow{2}{*}{$J^\mu$}&$0^{+-}:p^\mu$&$\times$&$\times$\\
			&$1^{--}:\epsilon^\mu$&&$-$\\ \hline
			\multirow{2}{*}{$\widetilde{J}^\mu$}&$0^{--}:p^\mu$&$\times$&$\times$\\ 
			&$1^{+-}:\epsilon^\mu$&&$-$\\ \hline
			\multirow{4}{*}{$J^{\{\mu\alpha\}}$}&$0^{++}:\eta^{\mu\alpha}$&$0$&\\
			&$0^{++}:p^\mu p^\alpha$&$\times$&\\
			&$1^{-+}:\epsilon^{\{\mu} p^{\alpha\}}$&$0$&\\ 
			&$2^{++}:\epsilon^{\mu\alpha}$&&\\ \hline
			\multirow{2}{*}{$J^{[\mu\alpha]}$}&$1^{-+}:\epsilon^{[\mu} p^{\alpha]}$&$0$&$\times$\\
			&$1^{++}:\epsilon^{\mu\alpha\rho\sigma} \epsilon_\rho p_\sigma$&&$-$\\ \hline 
			\multirow{4}{*}{$\widetilde{J}^{\{\mu\alpha\}}$}&$0^{--}:\eta^{\mu\alpha}$&$0$&\\
			&$0^{--}:p^\mu p^\alpha$&$\times$&\\
			&$1^{+-}:\epsilon^{\{\mu} p^{\alpha\}}$&$0$&\\ 
			&$2^{--}:\epsilon^{\mu\alpha}$&&\\ \hline
			\multirow{2}{*}{$\widetilde{J}^{[\mu\alpha]}$}&$1^{+-}:\epsilon^{[\mu} p^{\alpha]}$&$0$&$\times$\\
			&$1^{--}:\epsilon^{\mu\alpha\rho\sigma} \epsilon_\rho p_\sigma$&&$-$\\ \hline
			\multirow{8}{*}{$J^{\mu\nu\alpha}$}&$0^{-+}:\epsilon^{\mu\nu\alpha\rho}p_\rho$&&$-$\\
			&$0^{++}:\eta^{\mu\alpha}p^\nu - \eta^{\nu\alpha}p^\mu$&&$-$\\
			&$1^{-+}:\eta^{\mu\alpha}\epsilon^\nu - \eta^{\nu\alpha}\epsilon^\mu$&$0$&$-$\\
			&$1^{-+}:\epsilon^\mu p^\nu p^\alpha-\epsilon^\nu p^\mu p^\alpha$&$\times$&$-$\\
			&$1^{++}:\epsilon^{\mu\nu\rho\sigma} \epsilon_\rho p_\sigma p^\alpha$&$\times$&$-$\\ [2pt]
			&$1^{++}:{\epsilon^{\mu\alpha\rho\sigma} \epsilon_\rho p_\sigma p^\nu\atop-\{\mu\leftrightarrow\nu\}}$&$0$&$-$\\ [2pt]
			&$2^{++}:\epsilon^{\mu\alpha}p^\nu - \epsilon^{\nu\alpha}p^\mu$&&$-$\\
			&$2^{-+}:\epsilon^{\mu\nu\rho\sigma}{\epsilon_\rho}^\alpha p_\sigma$&&$-$
		\end{tabular}
	\end{ruledtabular}
\end{table}

For $J^{\mu\nu\alpha}$, the choice of polarization tensors is not unique. In four dimensions, the $\epsilon^{\mu\nu\rho\sigma}{\epsilon_\rho}^\alpha p_\sigma$ in Table~\ref{polar_summary} is equivalent to 
\begin{equation*}
	\epsilon^{\mu\alpha\rho\sigma}{\epsilon_\rho}^\nu p_\sigma-\{\mu\leftrightarrow\nu\},
\end{equation*}
while
\begin{equation*}
\epsilon^{\mu\nu\rho\sigma} \epsilon_\rho p_\sigma p^\alpha\quad \text{and}\quad \epsilon^{\mu\alpha\rho\sigma} \epsilon_\rho p_\sigma p^\nu-\{\mu\leftrightarrow\nu\}
\end{equation*} 
can be written as a linear combination of 
\begin{equation*}
p^2\epsilon^{\mu\nu\rho\sigma} \epsilon_\rho {\eta_\sigma}^\alpha\quad \text{and}\quad p^2\epsilon^{\mu\alpha\rho\sigma} \epsilon_\rho {\eta_\sigma}^\nu-\{\mu\leftrightarrow\nu\}.
\end{equation*}

It should be noted that different polarization tensors do not mean the corresponding states are independent; e.g., the Lorentz projector
\begin{equation}
	\sum_\lambda \epsilon_\lambda^{[\mu}p^{\alpha]}\,\epsilon_\lambda^{*\{\nu}p^{\beta\}}=\!\big(p^\alpha p^\beta g^{\mu\nu}-\{\mu\leftrightarrow\alpha\}\big) + \{\nu \leftrightarrow \beta\}
	\label{s_a_JP1}
\end{equation}
corresponds to the nonvanishing $J^{PC}=1^{-+}$ contribution in the correlator
\begin{equation}
	\langle T J^{[\mu\alpha]}(p)\,J^{\dagger\{\nu\beta\}}(p)\rangle.
\end{equation}
Here, the $[\mu\alpha]$ and $\{\nu\beta\}$ denote anti-symmetrization and symmetrization of indices, respectively. For simplicity, in this work we will not discuss the states corresponding to off-diagonal Lorentz projectors like Eq.~\eqref{s_a_JP1}.

For the numerical analysis, the following quark masses~\cite{pdg} and condensate values~\cite{qcd_sum_review} are used (both at $\mu=2\,\text{GeV}$):
\begin{align*}
	m_u &= 2.16\pm0.07\text{MeV},& m_d &= 4.70\pm0.07\text{MeV},\\
	m_s &= 93.5\pm 0.8\text{MeV},&\langle GG\rangle & = 0.07\pm0.02\,\text{GeV}^4,\\
	\langle \bar{q}q\rangle &= -(0.276)^3\,\text{GeV}^3,& \langle\bar{s}s\rangle &= 0.74\langle \bar{q}q\rangle,\\ 
	\langle\bar{q}Gq\rangle\! &= M_0^2\langle \bar{q}q\rangle,& \langle\bar{s}Gs\rangle &= M_0^2\langle \bar{s}s\rangle,\\
	M_0^2&=0.8\pm0.2\,\text{GeV}^2,&\langle GGG\rangle&=\!(8.2\pm\!2\,\text{GeV}^2) \langle GG\rangle,
	\label{qcd_parameters}
\end{align*}
where $q=u,d$; $\langle GG\rangle = \alpha_s\langle G^n_{\mu\nu}G^{n\,\mu\nu}\rangle$; $\langle GGG\rangle = \langle g^3 f^{abc}G^{a\ \nu}_\mu G^{b\ \rho}_\nu G^{c\ \mu}_\rho\rangle$; $\langle \bar{q}Gq\rangle=\langle\bar{q}gT^nG^n_{\mu\nu}\sigma^{\mu\nu}q\rangle$. The one-loop approximation of the running $\alpha_s$ is used:
\begin{equation}
	\alpha_s(\mu^2)=\frac{\alpha_s(m^2_\tau)}{1+\frac{\beta_0}{4\pi}\alpha_s(m^2_\tau)\log\big(\frac{\mu^2}{m^2_\tau}\big)},
\end{equation}
where~\cite{pdg}
\begin{equation*}
	m_\tau=1776.93\pm0.09\text{MeV},\quad \alpha_s(m^2_\tau)=0.314\pm0.014,
	\label{alpha_s}
\end{equation*}
and $\beta_0=9$ for $n_f=3$.

\begin{figure}[t!]
	\centering
	\includegraphics[width=8.4cm]{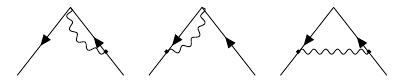}
	\caption{The diagrams involved in the renormalization of $\overline{\Psi}{\small\Gamma\overleftrightarrow{\nabla}}\Psi$ at one-loop level.}
	\label{ren_d0}
\end{figure}

For the calculation at next-to-leading order (NLO), the currents must be renormalized. The diagrams involved in the renormalization at the one-loop level are shown in Fig.~\ref{ren_d0}, where the fermion field $\Psi$ is renormalized implicitly, so that loops on the external legs need not be considered. The renormalized $\overline{\Psi}{\small\Gamma\overleftrightarrow{\nabla}}\Psi$ operator in the Feynman gauge and at $d=4-2\varepsilon$ can be written as:
\begin{equation}
	\begin{split}
		\Big(\overline{\Psi}\!_{f_a}\!\Gamma\overleftrightarrow{\nabla}\!\!_\alpha\Psi\!_{f_b}&\!\Big)_r=Z^{-1}_2\overline{\Psi}\!_{f_a}\!\Gamma\overleftrightarrow{\nabla}\!\!_\alpha\Psi\!_{f_b}\\
		&+\frac{g^2 C_F}{16\pi^2\varepsilon}\overline{\Psi}_{f_a}\big(\Gamma\slashed{\nabla}\gamma_\alpha-\gamma_\alpha\overleftarrow{\slashed{\nabla}}\Gamma \big)\Psi_{f_b}\\
		&-\frac{g^2C_F}{8\pi^2\varepsilon}i\overline{\Psi}_{f_a}\big(m_{f_b}\Gamma\gamma_\alpha+\gamma_\alpha\Gamma m_{f_a} \big)\Psi_{f_b}\\
		&+\!\frac{g^2C_F}{32\pi^2\varepsilon}i\overline{\Psi}\!_{f_a}\!\big(m\!_{f_a}\!\gamma^\mu\Gamma\gamma_\alpha\gamma_\mu\!+\!\gamma_\mu\gamma_\alpha\Gamma\gamma^\mu m\!_{f_b}\!\big)\!\Psi\!_{f_b}\\
		&-\frac{g^2C_F}{192\pi^2\varepsilon}\overline{\Psi}_{f_a}\gamma^\mu\gamma^\nu\Gamma\gamma_\nu\gamma_\mu\overleftrightarrow{\nabla}\!\!_\alpha\Psi_{f_b}\\
		&-\frac{g^2C_F}{48\pi^2\varepsilon}\overline{\Psi}_{f_a}\big(\gamma_\mu\gamma_\alpha\Gamma\slashed{\nabla}\gamma^\mu-\gamma^\mu\overleftarrow{\slashed{\nabla}}\Gamma\gamma_\alpha\gamma_\mu\big)\Psi_{f_b}\\
		&+\frac{g^2C_F}{96\pi^2\varepsilon}\overline{\Psi}_{f_a}\big(\gamma^\mu\slashed{\nabla}\Gamma\gamma_\alpha\gamma_\mu-\gamma_\mu\gamma_\alpha\Gamma\overleftarrow{\slashed{\nabla}}\gamma^\mu\big)\Psi_{f_b},
		\label{j_ren}
	\end{split}
\end{equation}
where $Z_2=1-\frac{g^2 C_F}{16\pi^2\varepsilon}$; the fields on the right-hand-side are bare fields; see Appendix~\ref{ren_} for more details.

A large number of diagrams are involved in the dimension-8 condensates. This is because the $\langle\bar{q}q\rangle\langle\bar{q}Gq\rangle$ contribution appears in the correlator at $O(\alpha_s)$ but is absent at $O(1)$, and the $\langle\bar{q}\nabla\nabla\nabla\nabla\nabla q\rangle$-type condensate also contributes to $\alpha_s\langle\bar{q}q\rangle\langle\bar{q}Gq\rangle$, as detailed in Appendix~\ref{d7_8_condensate}. The inclusion of dimension-8 condensates ensures the OPE is well converged, as shown in Appendix~\ref{ope_tau}.


\begin{figure*}[p]
	\centering
	\includegraphics[width=0.92\textwidth]{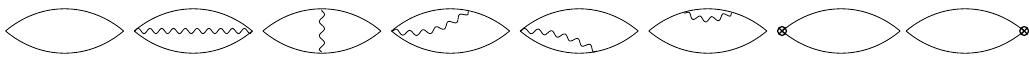}
	\caption{Diagrams for perturbative contributions. The last two diagrams are related to the counterterms.\label{ope_d0}}
\end{figure*}

\begin{figure*}[p]
	\centering
	\includegraphics[width=0.75\textwidth]{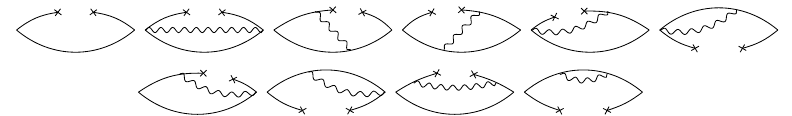}
	\caption{Diagrams for $m\langle\bar{q}q\rangle$ contributions.\label{ope_mqq}}
\end{figure*}

\begin{figure*}[p]
	\centering
	\includegraphics[width=0.92\textwidth]{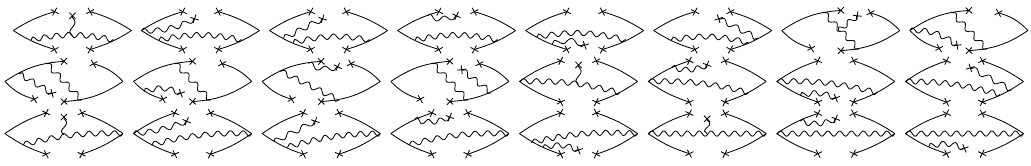}
	\includegraphics[width=0.6\textwidth]{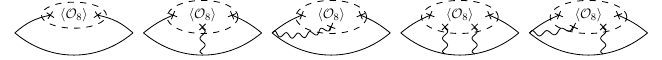}
	\caption{Diagrams for $\langle\bar{q}q\rangle\langle\bar{q}Gq\rangle$ and $m\langle GG\rangle\langle\bar{q}q\rangle$ contributions. The $\langle \mathcal{O}_8\rangle$ refers to condensates like $\langle\bar{q}\nabla\nabla\nabla\nabla\nabla q\rangle$, $\langle\bar{q}GG\nabla q\rangle$, $\langle\bar{q}DDDGq\rangle$, etc., which contribute to $\langle\bar{q}q\rangle\langle\bar{q}Gq\rangle$ and $m\langle GG\rangle\langle\bar{q}q\rangle$ via Eq.~\eqref{d8_expansion}.\label{ope_d8}}
\end{figure*}

\begin{table*}[p]
	\centering
	\caption{The corresponding diagrams for $\langle GG\rangle$, $m\langle\bar{q}Gq\rangle$, $\langle \bar{q}q\rangle^2$, and $m\langle \bar{q}q\rangle\langle GG\rangle$ contributions; Eq.~\eqref{d7_expansion} is used for the $m\langle \bar{q}q\rangle\langle GG\rangle$ contributions.\label{ope_digs}}
	\renewcommand{\arraystretch}{1.1}
	\begin{ruledtabular}
		\begin{tabular}{ll}
			&\multirow{3}{14cm}{{\raggedright\includegraphics[width=0.42\textwidth]{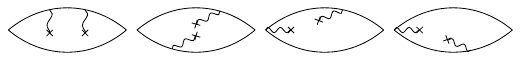}}}\\
			$\langle GG\rangle$& \\ 
			&\\ \hline
			&\multirow{3}{14cm}{{\raggedright\includegraphics[width=0.33\textwidth]{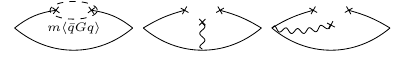}}}\\
			$m\langle\bar{q}Gq\rangle$&\\
			&\\ \hline
			&\multirow{3}{14cm}{{\raggedright\includegraphics[width=0.67\textwidth]{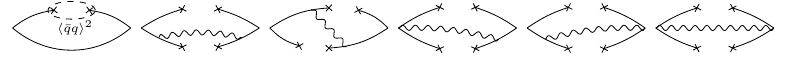}}}\\
			$\langle \bar{q}q\rangle^2$&\\ 
			&\\\hline
			&\multirow{3}{14cm}{{\raggedright\includegraphics[width=0.52\textwidth]{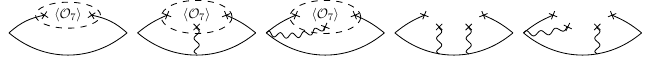}}}\\
			$m\langle \bar{q}q\rangle\langle GG\rangle$&\\
			&\\
		\end{tabular}
	\end{ruledtabular}
\end{table*}

\begin{table*}[p]
	\centering
	\caption{Diagrams and propagator notations for the $\langle G^3\rangle$ contribution. The use of combined propagators ensures the cancellation of infrared divergences, rendering each diagram infrared-free. For the detailed quark propagator structure, see Ref.~\cite{high_order_condensates}.\label{ope_ggg}}
	\renewcommand{\arraystretch}{1.1}
	\begin{ruledtabular}
		\begin{tabular}{ll}
			&\multirow{3}{14cm}{{\raggedright\includegraphics[width=0.7\textwidth]{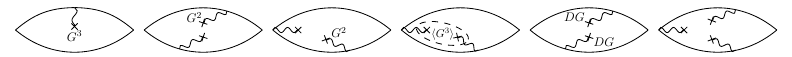}}}\\
			$\langle G^3\rangle$ diagrams&\\
			& \\ \hline
			&\multirow{5}{14cm}{{\raggedright\includegraphics[width=0.68\textwidth]{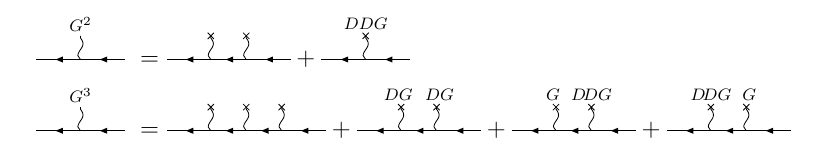}}}\\
			&\\
			The propagator&\\
			notations&\\
			&\\
		\end{tabular}
	\end{ruledtabular}
\end{table*}


%
The calculation results are listed in Appendix~\ref{corr_results}, while the involved diagrams are presented in Figs.~\ref{ope_d0}-\ref{ope_d8} as well as Tables.~\ref{ope_digs} and \ref{ope_ggg}; the perturbative and $m\langle\bar{q}q\rangle$ contributions are calculated to next-to-leading order (NLO). A Mathematica package~\cite{pack} was used for these evaluations.

\section{Numerical Analysis\label{num}}
\vspace{-0.25cm}
A preliminary investigation of the correlator is necessary before performing a numerical analysis. As summarized in Table~\ref{polar_summary}, some of the configurations have no $\langle GG\rangle$ contribution, while the $m\langle\bar{q}q\rangle$ contribution is nearly negligible for light quarks. In some other cases, the leading order (LO) perturbative contributions are absent. In both cases, the results are sensitive to the values of higher-dimensional condensates. Choosing the Lorentz projectors to be of dimension 2, e.g.,
\begin{equation}
	\Big(\frac{p^\mu p^\nu}{p^2}-g^{\mu\nu}\Big)\langle GG\rangle\rightarrow \big(p^\mu p^\nu-p^2g^{\mu\nu}\big)\frac{\langle GG\rangle}{p^2},
\end{equation}
then ensures that the $\langle GG\rangle$ contribution is involved in all cases except the conserved configurations. However, if a result is sensitive to the choice of the dimension of Lorentz projectors, it suggests that more corrections should be included to yield an accurate result. For simplicity, we choose all Lorentz projectors to be dimensionless in this work.

Therefore, we focus on the $J^P=2^\pm$ states corresponding to $J^{\{\mu\alpha\}}$ and $\widetilde{J}^{\{\mu\alpha\}}$ based on Table~\ref{polar_summary}; the masses for other cases are still presented if reasonable results can be derived. Focusing on the $J=2$ channel also offers a clear test of whether the $\overline{\Psi}{\small\Gamma \overleftrightarrow{\nabla}}\Psi$ operator couples to excited $q\bar{q}$ mesons, since the $q\bar{q}$ states with $J=2$ must be orbitally excited. In contrast, the $J^P=0^-$ and $1^-$ states can be mixtures of S-wave and excited states~\cite{3s-2d_mixing}, and in principle, the $\overline{\Psi}{\small\Gamma \overleftrightarrow{\nabla}}\Psi$ current couples to both ground and excited states if allowed by the quantum numbers.

\vspace{-0.35cm}
\subsection{Laplacian and Gaussian Sum Rules Analysis\label{sec_gsr}}
\vspace{-0.25cm}
The resonance information can be extracted from the correlator through the dispersion relation~\eqref{pole_continue}. However, a direct extraction is not feasible due to our limited knowledge of QCD at low energy. The Borel (Laplace) transformation is often used to estimate the lowest resonance mass. It is defined as~\cite{qsr,qsr_book,qsr_intro, qsr_intro_2,qsr_laplace, gaussian}
\begin{equation}
	\hat{\mathcal{B}}\,\Pi(q^2)=\lim_{\substack{-q^2,N\rightarrow\infty\\-q^2/N=M^2}}
	\frac{(-q^2)^{(N+1)}}{N!}\Big(\frac{d}{d q^2}\Big)^N \Pi(q^2),
	\label{borel_def}
\end{equation}
where $M^2$ is the Borel parameter, and one can find that
\begin{equation}
	\hat{\mathcal{B}}\,\frac{1}{s-q^2}=e^{-s/M^2}.
	\label{borel_1_sq2}
\end{equation}
Applying the Borel transformation to the dispersion relation then gives
\begin{equation}
		\hat{\mathcal{B}}\,\Pi(q^2) \!=\! \frac{1}{\pi}\,\hat{\mathcal{B}}\int_0^\infty\! ds\, \frac{\text{Im}\Pi(s)}{s-q^2}=\frac{1}{\pi}\int_0^\infty\! ds\, e^{-s/M^2}\, \text{Im}\Pi(s).
	\label{laplace_pi}
\end{equation}
According to local quark-hadron duality~\cite{local_q-h_duality, gaussian, gaussian_norm}, it is valid to truncate the integral at $s_0$ to isolate the resonance contribution. The mass of the resonance can then be easily derived from the ratio of moments:
\begin{equation}
	\mathcal{M}^n(M^2,s_0)=\frac{1}{\pi}\int_0^{s_0}ds\ s^n e^{- \frac{s}{M^2}}\, \text{Im}\Pi(s)=f^2 m^{2n}e^{-\frac{m^2}{M^2}}.
	\label{moment_qcd}
\end{equation}
By changing the integral kernel from $e^{-s/M^2}$ to $e^{-(s-t)^2/4\tau}$, the Laplace sum rule becomes the Gaussian sum rule (GSR)~\cite{gaussian, gaussian_norm, gsr_bayesian}; these two sum rules are actually intimately related. Note that the dispersion relation also gives~\cite{gaussian}
\begin{align}
	&\frac{\Pi(s\!+\!i Q^2)}{2i Q^2}-\frac{\Pi(s\!-\!i Q^2)}{2i Q^2}\nonumber\\
	&=\!\frac{1}{2iQ^2}\!\int_0^\infty\!\! dt\,\bigg[ \frac{1}{t\!-\!(s\!+\!iQ^2)}-\frac{1}{t\!-\!(s\!-\!iQ^2)}\bigg]\frac{1}{\pi}\text{Im}\Pi(t)\nonumber\\
	&=\!\!\int_0^\infty\!\! dt\, \frac{1}{(s\!-\!t)^2\!+\!(Q^2)^2}\frac{1}{\pi}\text{Im}\Pi(t),
	\label{disper_gauss}
\end{align}
where $s$ and $Q^2$ are real variables. According to Eqs.~\eqref{borel_def} and \eqref{borel_1_sq2}, replacing $-q^2\rightarrow Q^4$ and $M^2\rightarrow 4\tau$ in the Borel transformation and applying it to Eq.~\eqref{disper_gauss} then gives~\cite{gaussian,gsr_bayesian,1--_GSR_chen}
\begin{equation}
	\int_0^\infty \! dt\, e^{-\frac{(s-t)^2}{4\tau}}\frac{1}{\pi}\text{Im}\Pi(t).
	\label{cp2_gsr}
\end{equation}

The Gaussian kernel enables a more detailed examination of the spectral function. It makes it easy to probe different regions of the spectrum and study multiple resonances by varying the center of the kernel. Similar to Eq.~\eqref{moment_qcd}, one can isolate the resonance contribution as
\begin{subequations}
	\begin{equation}
		G(\tau,s,s_0) = \frac{1}{\sqrt{4\pi\tau}}\int_0^{s_0}dt\, e^{-\frac{(s-t)^2}{4\tau}}\frac{1}{\pi}\text{Im}\Pi(t).
		\label{gsr_pi}
	\end{equation}
On the hadronic side (``$\delta$ $+$ continuum'' ansatz), applying the Gaussian transformation yields
	\begin{equation}
		G_h(\tau,s)=\frac{f^2}{\sqrt{4\pi\tau}}e^{-\frac{(s-m^2)^2}{4\tau}}.
		\label{gsr_pole_s}
	\end{equation}
\end{subequations}
A numerical fitting procedure is often adopted in GSR, which reduces the bias from the choice of the continuum threshold $s_0$. The $G(\tau,s,s_0)$ and $G_h(\tau,s)$ here can be viewed as the functions of $s$ that describe the shape of the smeared resonance; on the hadron side, the narrow resonance $\delta(s-m^2)$ is smeared into a Gaussian peak $\exp[-\frac{(s-m^2)^2}{4\tau}]$. This is necessary since a comparison between the QCD side and the hadron side is valid only when averaged over a finite window (local quark-hadron duality~\cite{local_q-h_duality,gaussian,gaussian_norm}). In this work, we choose $\tau=10\,\text{GeV}^4$ as in Ref.~\cite{1--_GSR_chen}; the results are not sensitive to the choice of $\tau$~\cite{gsr_hybrid}, as shown in Appendix~\ref{ope_tau}. The renormalization-group improved GSR is obtained by setting $\mu^2=\sqrt{\tau}$~\cite{gaussian}. 

To make the mass estimation more robust, the $G(\tau,s,s_0)$ and $G_h(\tau,s)$ can be normalized as follows~\cite{gaussian_norm}:
\begin{subequations}
		\begin{align}
			\widetilde{G}(\tau,s,s_0)=&\,\,\frac{G(\tau,s,s_0)}{\int_{-\infty}^\infty ds\,G(\tau,s,s_0)},\label{gsr_norm_t}\\ \widetilde{G}_h(\tau,s)\,\,\,\,=&\,\,\frac{G_h(\tau,s)}{\int_{-\infty}^\infty ds\,G_h(\tau,s)} = \frac{1}{\sqrt{4\pi\tau}}e^{-\frac{(s-m^2)^2}{4\tau}},
			\label{gsr_norm_h}
		\end{align}
\end{subequations}
where the resonance strength $f$ has been eliminated, as it is not necessary for mass extraction. The mass $m$ and continuum threshold $s_0$ can then be estimated by minimizing the $\chi^2$:\vspace{-0.2cm}
\begin{equation}
	\chi^2(\tau,s_0)=\sum_{i=0}^N  \big[\widetilde{G}(\tau,s_i,s_0) - \widetilde{G}_h(\tau,s_i)\big]^2.\vspace{-0.35cm}
	\label{gsr_chi}
\end{equation}
Here,
\vspace{-0.25cm}
\begin{equation}
	N=\frac{s_\text{max} - s_\text{min}}{\delta s}+1,\quad s_i=s_\text{min}+i\, \delta s,\vspace{-0.1cm}
\end{equation}
and we choose:
\begin{equation*}
	s_\text{min}=-20\,\text{GeV}^2,\quad s_\text{max}=25\,\text{GeV}^2,\quad \delta s = 0.2\,\text{GeV}^2.
\end{equation*}

The values of $s_\text{min}$ and $s_\text{max}$ are chosen to be far from the peak of $\widetilde{G}(\tau,s,s_0)$, while $\delta s$ is chosen to be small enough to capture the detailed structure of $\widetilde{G}(\tau,s,s_0)$. The Monte Carlo method is also adopted to estimate the uncertainties. Based on the QCD parameters listed in Section~\ref{currents_basis}, we generate 2000 sets of input parameters by sampling from independent Gaussian distributions; the condensates $\langle \bar{q}q\rangle$ and $\langle \bar{s}s\rangle$ are assigned a 10\% relative uncertainty; 2000 independent GSR fittings are then performed for each configuration; see Appendix~\ref{_montecarlo} for more details.

\begin{figure*}[p]
	\centering
	\hspace{-0.08\textwidth}
	\begin{subfigure}{0.32\textwidth}
		\includegraphics[width=7.1cm, height=4cm]{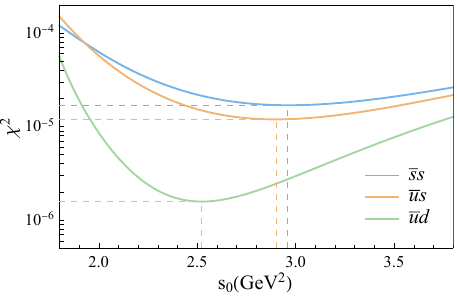}
		\vspace{-0.6cm}
		\caption{Minimized $\chi^2$ versus $s_0$.}
		\label{chi_2_s0}
	\end{subfigure}
	\hspace{0.14\textwidth}
	\begin{subfigure}{0.32\textwidth}
		\includegraphics[width=7.03cm, height=4cm]{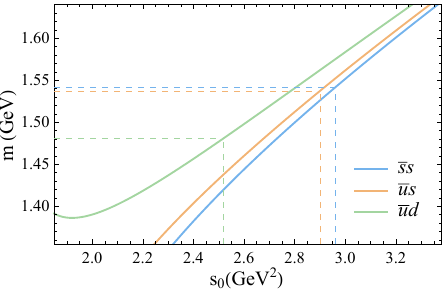}
		\vspace{-0.6cm}
		\caption{Fitted mass versus $s_0$. }
		\label{m_s0}
	\end{subfigure}
	\vspace{0.2cm}\\
	\hspace{-0.08\textwidth}
	\begin{subfigure}{0.32\textwidth}
		\includegraphics[width=7.23cm, height=4cm]{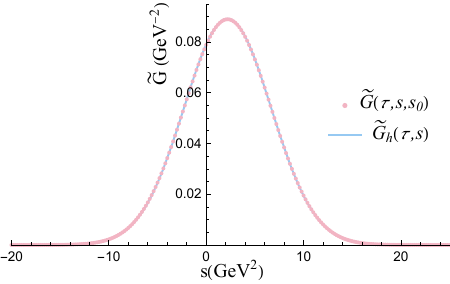}
		\vspace{-0.6cm}
		\caption{Best fit for the $\bar{u}d$ configuration. }
		\label{fit_ud}
	\end{subfigure}
	\hspace{0.14\textwidth}
	\begin{subfigure}{0.32\textwidth}
		\includegraphics[width=7.01cm, height=4cm]{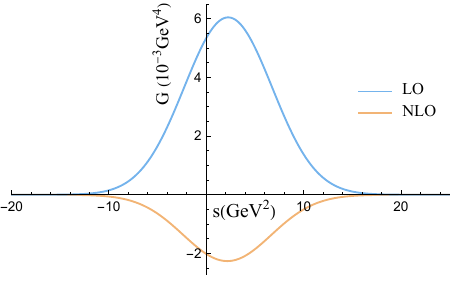}
		\vspace{-0.6cm}
		\caption{Perturbative contributions in GSR. }
		\label{lo_nlo}
	\end{subfigure}
	\caption{NLO GSR mass predictions for $2^{++}$ states extracted from $J^{\{\mu\alpha\}}$; the central values of the QCD parameters are used. Dashed lines in panels (a) and (b) mark the optimal value of $s_0$, as well as the corresponding $\chi^2$ and predicted mass. For $\bar{u}d$ configuration, the LO and NLO perturbative contributions after Gaussian transformation are shown in panel (d), where $s_0=3\,\text{GeV}^2$ is used; the NLO correction is considerable.}
	\label{GSR_J_s_2+}
\end{figure*}


\begin{table*}[p]
	\centering
	\caption{Summary of one-resonance GSR fitting results. The rows in bold font correspond to cases where $\langle GG\rangle$ and LO perturbative contributions are non-vanishing; for other rows, the accuracy is expected to be limited.\label{GSR_1-pole}}
	\renewcommand{\arraystretch}{1.3}
	\subfloat[LO GSR fitting results for the $2^-$ states. For the $2^{-+}$ channel, the predicted masses agree less well with the $2^{-+}$ states $\pi_2(1670)$, $K_2(1770)$, and $\eta_2(1870)$, compared to the NLO results in Table~\ref{gsr_NLO_1pole}.]{
		\begin{ruledtabular}
			\begin{tabular}{clcccccccc}
				&&\multicolumn{2}{c}{$\bar{u}d$}&&\multicolumn{2}{c}{$\bar{u}s
					$}&&\multicolumn{2}{c}{$\bar{s}s$}\vspace{-0.12cm}\\
				&& $m(\text{GeV})$&$s_0(\text{GeV}^2)$&&$m(\text{GeV})$&$s_0(\text{GeV}^2)$ &&$m(\text{GeV})$&$s_0(\text{GeV}^2)$ \\
				\hline
				$\boldsymbol{\widetilde{J}^{\{\mu\alpha\}}}$&$\boldsymbol{2^{--}:\epsilon^{\mu\alpha}}$&$\mathbf{1.50\pm0.09}$&$\mathbf{2.75\pm0.33}$&&$\mathbf{1.60\pm0.08}$&$\mathbf{3.11\pm0.30}$&&$\mathbf{1.59\pm0.08}$&$\mathbf{3.06\pm0.32}$\\ \hline
				$J^{\mu\nu\alpha}$&$2^{-+}:\epsilon^{\mu\nu\rho\sigma}{\epsilon_\rho}^\alpha p_\sigma$&$1.56\pm0.08$&$3.07\pm0.30$&&$1.61\pm0.07$&$3.24\pm0.30$&&$1.64\pm0.07$&$3.57\pm0.31$
			\end{tabular}
		\end{ruledtabular}
	}
	\vspace{0.2cm}
	\subfloat[Summary of NLO GSR fitting results. Some results (not shown) are not attainable or involve large relative errors. The degeneracy problem of the $2^{++}$ states corresponding to $J^{\{\mu\alpha\}}$ is solved in the two-resonance analysis shown in Table~\ref{GSR_2-pole}\label{gsr_NLO_1pole}.]{
		\begin{ruledtabular}
			\begin{tabular}{clcccccccc}
				&&\multicolumn{2}{c}{$\bar{u}d$}&&\multicolumn{2}{c}{$\bar{u}s$}&&\multicolumn{2}{c}{$\bar{s}s$}\vspace{-0.12cm}\\
				&& $m(\text{GeV})$&$s_0(\text{GeV}^2)$&&$m(\text{GeV})$&$s_0(\text{GeV}^2)$ &&$m(\text{GeV})$&$s_0(\text{GeV}^2)$ \\
				\hline
				$J^\mu$&$1^{--}:\epsilon^\mu$&$1.23\pm0.04$&$1.91\pm0.12$&&$1.33\pm0.04$&$2.20\pm0.14$&&$1.39\pm0.05$&$2.41\pm0.16$\\ \hline
				$\boldsymbol{J^{\{\mu\alpha\}}}$&$\boldsymbol{2^{++}:\epsilon^{\mu\alpha}}$&$\mathbf{1.53\pm0.18}$&$\mathbf{2.79\pm0.69}$&&$\mathbf{1.53\pm0.23}$&$\mathbf{2.84\pm0.87}$&&$\mathbf{1.53\pm0.25}$&$\mathbf{2.86\pm0.95}$\\ \hline
				$\boldsymbol{\widetilde{J}^{\{\mu\alpha\}}}$&$\boldsymbol{2^{--}:\epsilon^{\mu\alpha}}$&$\mathbf{1.74\pm0.10}$&$\mathbf{3.64\pm0.41}$&&$\mathbf{1.83\pm0.08}$&$\mathbf{4.05\pm0.38}$&&$\mathbf{1.89\pm0.08}$&$\mathbf{4.28\pm0.37}$\\ \hline
				$\widetilde{J}^{[\mu\alpha]}$&$1^{--}:\epsilon^{\mu\alpha\rho\sigma} \epsilon_\rho p_\sigma$&$1.14\pm0.04$&$1.66\pm0.12$&&$1.21\pm0.05$&$1.85\pm0.16$&&$1.27\pm0.05$&$2.02\pm0.17$\\ \hline
				\multirow{4}{*}{$J^{\mu\nu\alpha}$}&$0^{-+}:\epsilon^{\mu\nu\alpha\rho}p_\rho$&$1.09\pm0.04$&$1.54\pm0.13$&&$1.10\pm0.05$&$1.53\pm0.14$&&$1.08\pm0.06$&$1.46\pm0.16$\\
				&$2^{++}:\epsilon^{\mu\alpha}p^\nu \!- \epsilon^{\nu\alpha}p^\mu$&$1.42\pm0.27$&$2.56\pm0.82$&&$1.32\pm0.26$&$2.30\pm0.72$&&$1.28\pm0.26$&$2.20\pm0.70$\\
				&$2^{-+}:\epsilon^{\mu\nu\rho\sigma}{\epsilon_\rho}^\alpha p_\sigma$&$1.74\pm0.08$&$3.77\pm0.36$&&$1.83\pm0.07$&$4.14\pm0.35$&&$1.89\pm0.07$&$4.39\pm0.36$
			\end{tabular}
		\end{ruledtabular}
	}
\end{table*}

To visualize the solution's dependence on the continuum threshold $s_0$, we first present the GSR results where $s_0$ is fixed to different values and only $m$ is treated as a fitting parameter. As shown in Fig.~\ref{GSR_J_s_2+}, the $J^{\{\mu\alpha\}}$ yields
$J^{PC}=2^{++}$ state masses $\simeq1.48\,\text{GeV}$ for $\bar{u}d$ configuration, while for $\bar{u}s$ and $\bar{s}s$ configurations, the derived masses are nearly degenerate ($\simeq1.54\,\text{GeV}$). Additionally, the NLO perturbative contribution is around $-40\%$ of LO perturbative contribution, indicating the necessity of NLO corrections; the masses of $J^P=2^-$ states in Table~\ref{GSR_1-pole} also confirms this necessity.

Compared with the $2^{++}$ excited $q\bar{q}$ candidates $a_2(1320)$, $K_2^*(1430)$, and $f_2^\prime(1525)$, the estimated masses for $\bar{u}d$ and $\bar{u}s$ do not agree very well. After applying Monte Carlo uncertainty analysis, the obtained masses are degenerate as shown in Table~\ref{GSR_1-pole}. Nevertheless, the continuum thresholds in Table~\ref{GSR_1-pole} and the masses in Fig.~\ref{GSR_J_s_2+} still imply that a mass hierarchy $\bar{u}d<\bar{u}s<\bar{s}s$ should exist. As shown later in Section.~\ref{gsr_2_poles}, this inconsistency is clarified when a second resonance is included.

For the $J^P=2^-$ channel, the derived masses for $C$-even and -odd states are nearly identical, as shown in Table~\ref{GSR_1-pole}. The masses for $\bar{u}d$, $\bar{u}s$, and $\bar{s}s$ configurations agree well with the masses of $2^{-+}$ states $\pi_2(1670)$, $K_2(1770)$, and $\eta_2(1870)$.

For correlators that have no $\langle GG\rangle$ contribution in the imaginary part or have no LO perturbative contribution, the accuracy of these results may be limited as mentioned previously. The $1^{--}$ states masses given by $J^\mu$ are around $1.2-1.4\,\text{GeV}$, which are close to the masses of $\rho(1450)$, $K^*(1410)$, and $\phi(1680)$; the $0^{-+}$ states masses given by $J^{\mu\nu\alpha}$ are around $1.1\,\text{GeV}$, which are close to the masses of $\pi(1300)$, $K(1460)$, and $\eta(1475)$. These predictions are much heavier than the ground $0^{-+}$ and $1^{--}$ states, implying that the $\overline{\Psi}\Gamma{\small\overleftrightarrow{\nabla}}\Psi$ operators should couple preferentially to the excited $q\bar{q}$ mesons in these channels.

For the $1^-$ and $0^+$ channels, the LSR results agree better with experiments compared to those from GSR. However, quantitative error estimation and the extraction of multiple resonances are difficult in LSR; we therefore provide these results in Appendix \ref{sec_lsr} as a supplement and present only the GSR results in the main text.


\subsection{Two-Resonance Gaussian Sum Rules Analysis\label{gsr_2_poles}}

Compared with the $2^{-\pm}$ channels, the predicted masses for the $2^{++}$ states show poor agreement with experiment, and the errors are also larger. Since two $2^{++}$ nonets are observed experimentally~\cite{pdg}, one possible reason is that the second resonance prevents the isolation of the lowest-lying resonance in the previous analysis. Here, we present a two-resonance analysis for the $2^{++}$ states. By modifying the spectral density to
\begin{equation}
	\rho(t)=f^2_1 \delta(t-m_1^2) + f^2_2 \delta(t-m_2^2) +\theta(t-s_0)\rho(t),
\end{equation}
the normalized GSR in Eq.~\eqref{gsr_norm_h} then becomes
\begin{equation}
	\widetilde{G}(\tau,s)=\frac{1}{\sqrt{4\pi\tau}}\Big[r \,e^{-\frac{(s-m_1^2)^2}{4\tau}} + (1-r)\,e^{-\frac{(s-m_2^2)^2}{4\tau}}\Big],
	\label{2-deltas}
\end{equation}
where $r$ is the relative coupling strength
\begin{equation}
	r=\frac{f^2_1}{f^2_1+f^2_2}.
\end{equation}

\begin{figure*}[t]
	\raggedright 
	\hspace{0cm}
	\includegraphics[width=5.6cm, height=3.68cm]{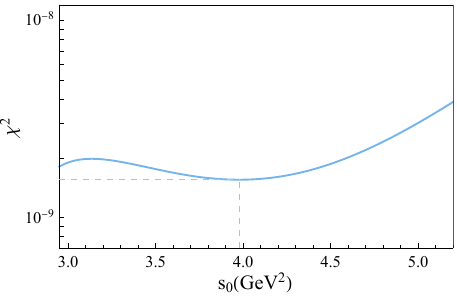}
	\hspace{0.35cm}
	\includegraphics[width=5.4cm, height=3.68cm]{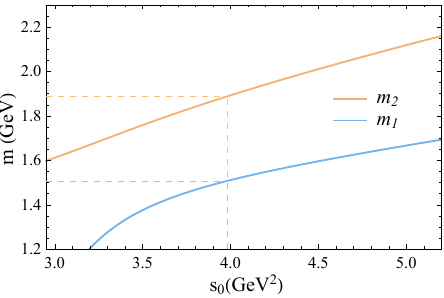}
	\hspace{0.35cm}
	\includegraphics[width=5.2cm, height=3.68cm]{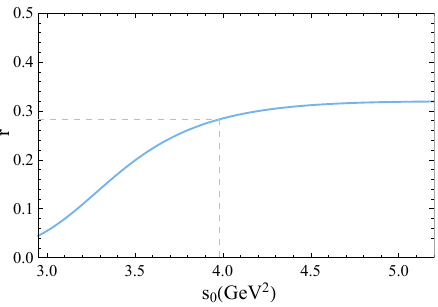}
	\caption{The two-resonance GSR fitting results of the $2^{++}$ $\bar{u}d$ state corresponding to $J^{\{\mu\alpha\}}$. Dashed lines mark the optimal value of $s_0$, as well as the corresponding $\chi^2$, masses, and relative coupling strength $r$.}
	\label{GSR_ud_2+}
\end{figure*}

\begingroup
\thinmuskip=0mu
\medmuskip=1mu
\thickmuskip=1mu
\begin{table*}[t]
	\centering
	\caption{Summary of the two-Resonance GSR fitting results for different $J^{PC}=2^{++}$ configurations. \label{GSR_2-pole}}
	\renewcommand{\arraystretch}{1.45}
	\subfloat[$m_1$, $r$, and $s_0$; obtained by fixing $m_2$ to the masses of $a_2(1700)$, $K_2^*(1980)$, and $f_2(1950)$~\cite{pdg} for $\bar{u}d$, $\bar{u}s$, and $\bar{s}s$, respectively. The mass of $K_2^*(1980)$, $1990_{-50}^{+60}\,\text{MeV}$, is manually set to be $1990\pm55\,\text{MeV}$ in the Monte Carlo uncertainty analysis.]{\resizebox{18.1cm}{!}{
			\begin{ruledtabular}
				\begin{tabular}{cl*{13}{c}}
					&&\multicolumn{3}{c}{$\bar{u}d$}&&&\multicolumn{3}{c}{$\bar{u}s$}&&&\multicolumn{3}{c}{$\bar{s}s$}\vspace{-0.08cm}\\ 
					&&$m_1(\text{GeV})$&$r$ &$s_0(\text{GeV}^2)$&&&$m_1(\text{GeV})$&$r$ &$s_0(\text{GeV}^2)$&&&$m_1(\text{GeV})$&$r$ &$s_0(\text{GeV}^2)$\\
					\hline
					$\medmath{\boldsymbol{J^{\{\mu\alpha\}}}}$&$\medmath{\boldsymbol{2^{++}\!\!\!:\!\epsilon^{\mu\alpha}}}$&$\medmath{\mathbf{1.23\pm0.15}}$&$\medmath{\mathbf{0.17\pm0.12}}$&$\medmath{\mathbf{3.32\pm0.09}}$&&&$\medmath{\mathbf{1.50\pm0.10}}$&$\medmath{\mathbf{0.26\pm0.06}}$&$\medmath{\mathbf{4.42\pm0.25}}$&&&$\medmath{\mathbf{1.44\pm0.08}}$&$\medmath{\mathbf{0.25\pm0.05}}$&$\medmath{\mathbf{4.25\pm0.08}}$\\ \hline
					$\medmath{J^{\mu\nu\alpha}}$&$\medmath{2^{++}\!\!\!:{\epsilon^{\mu\alpha}p^\nu \atop- \epsilon^{\nu\alpha}p^\mu}}$&$\medmath{1.24\pm0.10}$&$\medmath{0.29\pm0.07}$&$\medmath{3.29\pm0.08}$&&&$\medmath{1.47\pm0.09}$&$\medmath{0.32\pm0.06}$&$\medmath{4.43\pm0.25}$&&&$\medmath{1.43\pm0.08}$&$\medmath{0.33\pm0.07}$&$\medmath{4.20\pm0.11}$
				\end{tabular}
			\end{ruledtabular}
		}
	}
	\vspace{12pt}
	\subfloat[$m_2$, $r$, and $s_0$; obtained by fixing $m_1$ to the masses of $a_2(1320)$, $K_2^*(1430)$, and $f_2^\prime(1525)$~\cite{pdg} for $\bar{u}d$, $\bar{u}s$, and $\bar{s}s$, respectively.]{\resizebox{18.1cm}{!}{
			\begin{ruledtabular}
				\begin{tabular}{cl*{13}{c}}
					&&\multicolumn{3}{c}{$\bar{u}d$}&&&\multicolumn{3}{c}{$\bar{u}s$}&&&\multicolumn{3}{c}{$\bar{s}s$}\vspace{-0.08cm}\\
					&&$m_2(\text{GeV})$&$r$ &$s_0(\text{GeV}^2)$&&&$m_2(\text{GeV})$&$r$ &$s_0(\text{GeV}^2)$&&&$m_2(\text{GeV})$&$r$ &$s_0(\text{GeV}^2)$\\
					\hline
					$\medmath{\boldsymbol{J^{\{\mu\alpha\}}}}$&$\medmath{\boldsymbol{2^{++}\!\!\!:\!\epsilon^{\mu\alpha}}}$&$\medmath{\mathbf{1.76\pm0.08}}$&$\medmath{\mathbf{0.17\pm0.05}}$&$\medmath{\mathbf{3.54\pm0.32}}$&&&$\medmath{\mathbf{1.90\pm0.10}}$&$\medmath{\mathbf{0.22\pm0.04}}$&$\medmath{\mathbf{4.10\pm0.43}}$&&&$\medmath{\mathbf{2.00\pm0.10}}$&$\medmath{\mathbf{0.26\pm0.05}}$&$\medmath{\mathbf{4.51\pm0.48}}$\\ \hline
					$\medmath{J^{\mu\nu\alpha}}$	&$\medmath{2^{++}\!\!\!:{\!\epsilon^{\mu\alpha}p^\nu \atop- \epsilon^{\nu\alpha}p^\mu}}$&$\medmath{1.71\pm0.24}$&$\medmath{0.22\pm0.16}$&$\medmath{3.38\pm0.90}$&&&$\medmath{1.89\pm0.17}$&$\medmath{0.30\pm0.11}$&$\medmath{4.05\pm0.77}$&&&$\medmath{2.01\pm0.13}$&$\medmath{0.33\pm0.11}$&$\medmath{4.56\pm0.67}$
				\end{tabular}
			\end{ruledtabular}
		}
	}
\end{table*}
\endgroup
Following the same procedure as in Eq.~\eqref{gsr_chi}, the parameters $s_0$, $m_1$, $m_2$, and $r$ can be obtained. As shown in Fig.~\ref{GSR_ud_2+}, we first present the investigation where $s_0$ is fixed to different values. The fitted masses $m_1\simeq1.5\,\text{GeV}$ and $m_2\simeq 1.9\,\text{GeV}$ are compatible with the masses of $a_2(1320)$ and $a_2(1700)$; in particular, the mass difference $m_2-m_1\simeq0.4\,\text{GeV}$ agrees precisely. The relative coupling strength $r\simeq 0.3$ also indicates that a single-resonance scenario is inadequate. However, the precision of evaluated correlator is limited, especially in the low-energy region; the fitting procedure involving four parameters is delicate and prone to failure, and the solution is sensitive to the input QCD parameters.

To resolve this, a procedure similar to that in Ref.~\cite{1--_GSR_chen} is adopted: by fixing the mass of one resonance to the value observed in experiments, the $s_0$, $r$, and the mass of the other state are then determined by the fitting procedure. As summarized in Table~\ref{GSR_2-pole}, all predicted masses agree with experiments within errors, where the incorrect mass hierarchy of $m_1$ for $\bar{u}s$ and $\bar{s}s$ states is caused by fixing $m_2$ to the masses of $K_2^*(1980)$ and $f_2(1950)$, respectively; a heavier $m_1$ is then preferred for the $\bar{u}s$ state in fitting procedure. Additionally, the parameter $r$ consistently shows that $\widetilde{J}^{\{\mu\alpha\}}$ couples preferentially to the heavier $2^{++}$ states, which explains the poor agreement with experiments in the previous one-resonance analysis.

\section{Discussion and Conclusion\label{conclusion}}

Using the $\overline{\Psi}{\small\Gamma{\small\overleftrightarrow{\nabla}}}\Psi$-type operator, we obtain several $J^P=2^\pm$ nonets with masses in good agreement with experiments; the derived masses for $J=0,1$ states are also compatible with experiments as shown in Table~\ref{GSR_1-pole} and Appendix~\ref{sec_lsr}, though more work is needed to yield more accurate results. On the other hand, many alternative operator constructions exist. In Ref.~\cite{PhysRevD.79.074025}, the operator $\overline{\Psi}\Gamma(z\cdot{\small\overleftrightarrow{\nabla}})^2\Psi$ was introduced to study the $p$-wave state, where $z^2=-1$ for the purpose of probing the wave function in a space-like direction. Such a deliberate construction may enable a more precise selection of the states of interest.

The detailed NLO GSR analysis shows that for the $J^P=2^{-+}$ states corresponding to $J^{\mu\nu\alpha}$, the predicted masses for the $\bar{u}d$, $\bar{u}s$, and $\bar{s}s$ configurations are $1.74\pm0.08\,\text{GeV}$, $1.83\pm0.07\,\text{GeV}$, and $1.89\pm0.07\,\text{GeV}$, respectively. These results align remarkably well with the $\pi_2(1670)$, $K_2(1770)$, and $\eta_2(1870)$, respectively, indicating the validity of the $\overline{\Psi}\Gamma{\small\overleftrightarrow{\nabla}}\Psi$ operator methodology in this channel.

The analysis of $J^P=2^{++}$ states corresponding to the $J^{\{\mu\alpha\}}$ and $J^{\mu\nu\alpha}$ current is more intricate. The single-resonance GSR analysis is unable to yield results that agree well with experiment, and a two-resonance GSR analysis was then performed. By fixing the mass of one resonance to its known experimental value, the mass of the other resonance is successfully extracted. The obtained masses agree well with the experimental candidates; one is the nonet containing $a_2(1320)$, $K_2^*(1430)$, and $f_2^\prime(1525)$; the other is the nonet containing $a_2(1700)$, $K_2^*(1980)$, and $f_2(1950)$. Furthermore, the relative coupling strength $r$ indicates that $J^{\{\mu\alpha\}}$ couples preferentially to the heavier $2^{++}$ resonances.

In conclusion, this work demonstrates that $\overline{\Psi}\Gamma{\small\overleftrightarrow{\nabla}^n}\Psi$-type operators are efficient and versatile tools for the investigation of excited $q\bar{q}$ mesons. This encourages studies of excited hadrons beyond $q\bar{q}$ mesons.

\appendix

\section{Laplace Sum Rules Analysis\label{sec_lsr}}

To estimate the resonance mass, the Laplace sum rule (LSR) is a robust and widely used method. However, it is difficult to handle two-resonance extraction compared to GSR; therefore, we only present the LSR results here for several $1^-$ and $0^+$ channels as supplements. For these states, the LSR yields results that agree better with experimental data compared to GSR.

The mass of lowest resonance can be derived from the ratio of moments in Eq.~\eqref{moment_qcd}
\begin{equation}
	\mathcal{R}^n(M^2,s_0)=\frac{\mathcal{M}^{n+1}(M^2,s_0)}{\mathcal{M}^n(M^2,s_0)}=m^2.
	\label{ratio}
\end{equation}
The parameters $M^2$ and $s_0$ are commonly constrained within the Borel Window~\cite{qcd_book}. One constraint is $\mathcal{M}^0(M^2,s_0)/\mathcal{M}^0(M^2,\infty)\geq50\%$, so that the resonance contribution dominates; another is that the contribution received from the highest-dimensional condensate is less than $10\%$, to ensure the OPE converges. The stability criterion~\cite{mini_qcd} is commonly adopted as a compensation for determining the mass, which involves identifying the extremum or the most stable region in the curve of $\sqrt{\mathcal{R}^n}(M^2)$.

While this method is simple and convenient, it presents difficulties for error estimation. One may also adopt a numerical fitting procedure for Eq.~\eqref{moment_qcd}. However, we found that the $\chi^2$ fitting, defined by
\begin{equation}
	\chi^2(f,\!m,\!s_0)\!=\!\!\sum_i\!\bigg[f^2 e^{-\!\frac{m^2}{M_i^2}}\!-\frac{1}{\pi}\!\!\int_0^{s_0}\!\!ds\, e^{-\! \frac{s}{M_i^2}} \text{Im}\Pi(s)\bigg]^2
\end{equation}
is sensitive to the choice of the fitting range $[M^2_{\text{min}}, M^2_{\text{max}}]$, unlike Eq.~\eqref{gsr_chi}, which is not sensitive to the choice of $s_{\text{min}}$ and $s_{\text{max}}$ provided they are away from the peak. Therefore, we adopt only the ratio method in Eq.~\eqref{ratio} to extract the masses, and list only the approximate values without specific uncertainties in Table~\ref{LSR_masses}.

\begin{figure*}[ht!]
	\centering
	\begin{subfigure}{0.4\textwidth}
		\includegraphics[width=7.2cm, height=4cm]{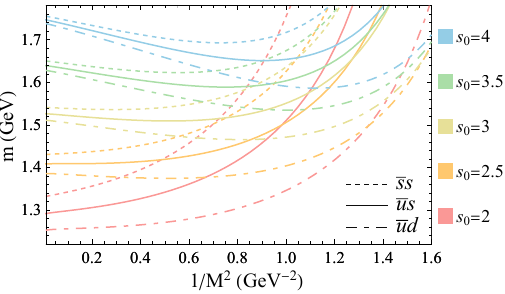}
		\vspace{-0.6cm}
		\caption{$1^{--}$ states given by $J^\mu$.}
		\label{J_1-}
	\end{subfigure}
	\hspace{2cm}
	\begin{subfigure}{0.4\textwidth}
		\includegraphics[width=7.2cm, height=4cm]{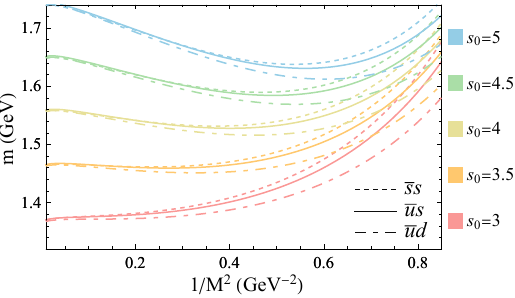}
		\vspace{-0.6cm}
		\caption{$1^{-+}$ states given by $J^{\{\mu\alpha\}}$. }
		\label{J_s_1-}
	\end{subfigure}\\
	\vspace{0.35cm}
	\begin{subfigure}{0.4\textwidth}
		\includegraphics[width=7.2cm, height=4cm]{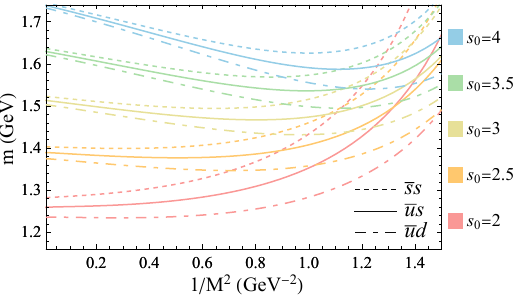}
		\vspace{-0.6cm}
		\caption{$1^{--}$ states given by $\widetilde{J}^{[\mu\alpha]}$. }
		\label{J_a5_1-}
	\end{subfigure}
	\hspace{2cm}
	\begin{subfigure}{0.4\textwidth}
		\includegraphics[width=7.2cm, height=4cm]{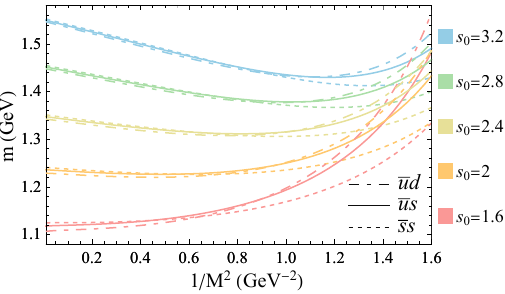}
		\vspace{-0.6cm}
		\caption{$0^{-+}$ states given by $J^{\mu\nu\alpha}$. }
		\label{J_0-}
	\end{subfigure}
	\vspace{0cm}
	\caption{NLO LSR mass predictions: $\sqrt{\mathcal{R}^0}(\text{GeV})$ versus $1/M^2\,(\text{GeV}^{-2})$ for different $s_0(\text{GeV}^2)$. The Corresponding polarization tensor are listed in Table~\ref{polar_summary}.}
	\label{LSR_J_01_m}
\end{figure*}

\begin{figure*}[ht!]
	\centering
	\includegraphics[width=7.34cm, height=4cm]{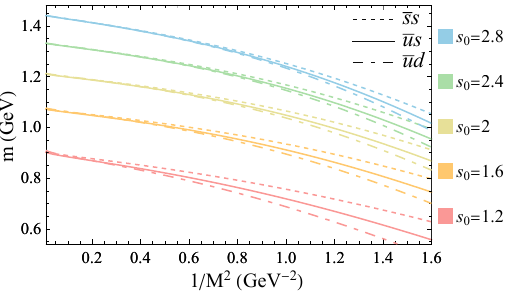}
	\hspace{1.86cm}
	\includegraphics[width=7.2cm, height=4cm]{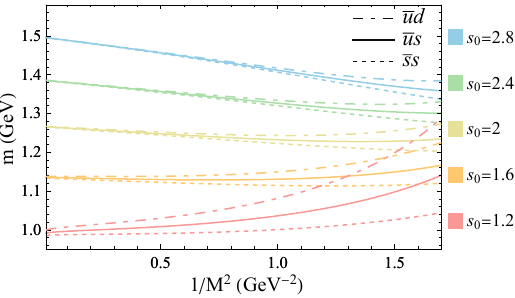}
	\caption{NLO LSR mass predictions for $0^{++}$ nonets given by $J^{\mu\nu\alpha}$:  $\sqrt{\mathcal{R}^n}$ versus $1/M^2\,(\text{GeV}^{-2})$ for different $s_0(\text{GeV}^2)$. The left and right panels corresponding to using $\sqrt{\mathcal{R}^0}$ and $\sqrt{\mathcal{R}^1}$, respectively; note the latter shows inverse hierarchy $\bar{u}d>\bar{u}s>\bar{s}s$.}
	\label{LSR_J_0_p}
\end{figure*}

Additionally, we list the working ranges of $M^2$ and $s_0$ (constrained by the Borel window) in Table~\ref{LSR_borel}. Note that these two parameters are not strictly fixed by the Borel window; the $s_0$ in Table~\ref{LSR_borel} are chosen to ensure $M^2$ and $s_0$ are compatible with the stability criteria. As shown in Figs.~\ref{LSR_J_01_m} and \ref{LSR_J_0_p}, the curves are stable around these ranges, except for the $1^{--}$ states given by $J^\mu$.

As shown in Fig.~\ref{LSR_J_01_m}, the stability criterion yields $1^{-\pm}$ state masses around $1.4-1.6\,\text{GeV}$ for $J^\mu$, $J^{\{\mu\alpha\}}$, and $\widetilde{J}^{[\mu\alpha]}$, while the $0^{-+}$ states corresponding to $J^{\mu\nu\alpha}$ have masses around $1.2\,\text{GeV}$. The masses of $1^{--}$ states agree well with radially excited $1^{--}$ nonet $\rho(1450)$, $K^*(1410)$, and $\phi(1680)$; similarly, the $0^{-+}$ states agree well with radially excited $0^{-+}$ nonet $\pi(1300)$, $K(1460)$, and $\eta(1475)$.

Interestingly, the $1^{-+}$ states, which cannot be $q\bar{q}$ mesons, have masses around 1.5 GeV, as shown in Fig.~\ref{J_s_1-}. This result agrees well with the exotic state $\pi_1(1600)$~\cite{pdg}, which is often interpreted as a hybrid state ($q\bar{q}$ with an excited/constituent gluon).

In contrast, for $0^{++}$ states corresponding to $J^{\mu\nu\alpha}$, the $\sqrt{\mathcal{R}^0}(M^2)$ shows no plateau, while the $\sqrt{\mathcal{R}^1}(M^2)$ still yields mass prediction around $1.1-1.2\,\text{GeV}$, as shown in Fig.~\ref{LSR_J_01_m}. The inverse hierarchy $\bar{u}d>\bar{u}s>\bar{s}s$ agrees with the masses of $a_0(1450)$ and $K_0^*(1430)$, which have $I=1$ and $1/2$, respectively, while the $I=0$ states remain undetermined.

\begin{table}[t!]
	\centering
	\caption{Summary of LSR results for several $J^P=0^\pm$ and $1^\pm$ states. \label{LSR_}}%
	\renewcommand{\arraystretch}{1.4}
	\subfloat[Mass estimates (in GeV).\label{LSR_masses}]{
		\begin{ruledtabular}
			\begin{tabular}{cclcccccc}
				&&&$\bar{u}d$&&$\bar{u}s$&&$\bar{s}s$&\\
				\hline
				&$J^\mu$&$1^{--}:\epsilon^\mu$&$\sim1.4$&&$\sim1.5$&&$\sim1.6$&\\ \hline
				&$J^{\{\mu\alpha\}}$&$1^{-+}:\epsilon^{\{\mu}p^{\alpha\}}$&$\sim1.5$&&$\sim1.5$&&$\sim1.5$&\\ \hline
				&$\widetilde{J}^{[\mu\alpha]}$&$1^{--}:\epsilon^{\mu\alpha\rho\sigma}\epsilon_\rho p_\sigma$&$\sim1.3$&&$\sim1.4$&&$\sim1.5$&\\ \hline
				&\multirow{2}{*}{$J^{\mu\nu\alpha}$}&$0^{-+}:\epsilon^{\mu\nu\alpha\rho} p_\rho$&$\sim1.3$&&$\sim1.3$&&$\sim1.3$&\\
				&&$0^{++}:\eta^{\mu\alpha} p^\nu-\eta^{\nu\alpha} p^\mu$&$\sim1.2$&&$\sim1.1$&&$\sim1.1$&
			\end{tabular}
		\end{ruledtabular}
	}
	\vspace{12pt}
	\subfloat[Ranges of $1/M^2$ (in $\text{GeV}^{-2}$) constrained by the Borel window for specific values of $s_0$ (in $\text{GeV}^2$); polarization tensors for each row are the same as in Table~\ref{LSR_masses}.\label{LSR_borel}]{
		\begin{ruledtabular}
			\begin{tabular}{cccccccc}
				&&$\bar{u}d$&&$\bar{u}s$&&$\bar{s}s$&\\
				\hline
				$J^\mu(1^{--};s_0=4)$&&$0.7\!-\!1.54$&&$0.72\!-\!1.53$&&$0.74\!-\!1.51$&\\ \hline
				$J^{\{\mu\alpha\}}(1^{-+};s_0=4)$&&$0.43\!-\!0.67$&&$0.43\!-\!0.73$&&$0.43\!-\!0.8$&\\ \hline
				$\widetilde{J}^{[\mu\alpha]}(1^{--};s_0=3)$&&$0.92\!-\!1.27$&&$0.95\!-\!1.32$&&$0.98\!-\!1.36$&\\ \hline
				$J^{\mu\nu\alpha}(0^{-+};s_0=2.8)$&&$0.93\!-\!1.32$&&$0.94\!-\!1.32$&&$0.95\!-\!1.51$&\\
				$J^{\mu\nu\alpha}(0^{++};s_0=2.4)$&&$1.02\!-\!1.36$&&$1.02\!-\!1.45$&&$1.03\!-\!1.66$&
			\end{tabular}
		\end{ruledtabular}
	}
\end{table}

\break
\section{Dimension-7 and -8 Condensates\label{d7_8_condensate}}

For the calculation of $m\langle GG\rangle\langle \bar{q}q\rangle$ and $\langle\bar{q}q\rangle\langle\bar{q}Gq\rangle$ contributions, the following equations are used:


\begin{widetext}
	\begin{align}
		&\langle \bar{q}_i\nabla^\mu\nabla^\nu\nabla^\alpha\nabla^\beta q_j\rangle=\nonumber\\
		&\frac{1}{576} \Big[-\gamma ^{\alpha \nu } g^{\beta \mu }\left(Q_1^7-2 Q_2^7+Q_4^7+m Q^6-2 m^2 Q^5\right) - \left(\gamma
		^{\beta \nu } g^{\alpha \mu }+\gamma ^{\alpha \mu } g^{\beta \nu }\right)\left(2 Q_1^7-Q_2^7-3 Q_3^7+Q_4^7-m Q^6-m^2 Q^5\right)\nonumber\\[0pt]
		&+\left(\gamma ^{\mu \nu } g^{\alpha \beta
		} \!+\!\gamma ^{\alpha \beta } g^{\mu \nu }\right)\left(Q_1^7\!-\!2 Q_2^7\!-\!3 Q_3^7\!+\!m Q^6\!-\!2 m^2 Q^5\right)  \!+\!  g^{\alpha \nu } g^{\beta \mu }\left(12 Q_1^7\!-\!3 Q_2^7\!-\!6 Q_3^7\!+\!4 Q_4^7\!-\!8 m Q^6\!-\!6 m^2 Q^5\!+\!6 m^4 Q^3\right) \nonumber\\[0pt]
		&+
		g^{\alpha \mu } g^{\beta \nu }\left(6 Q_1^7\!-\!3 Q_2^7\!-\!6 Q_3^7\!+\!4 Q_4^7\!-\!8 m Q^6\!-\!6 m^2 Q^5\!+\!6 m^4 Q^3\right)\!-\! g^{\alpha \beta } g^{\mu \nu }\left(3 Q_2^7\!+\!6 Q_3^7\!+\!2
		Q_4^7\!-\!4 m Q^6\!+\!6 m^2 Q^5\!-\!6 m^4 Q^3\right)\nonumber\\[0pt]
		&-\gamma ^{\beta \mu } g^{\alpha \nu }\left(3 Q_1^7+Q_4^7-3 m Q^6\right) -3 \gamma ^{\alpha \beta \mu \nu } Q_2^7\Big]_{ji};\label{d7_expansion}
	\end{align}
	\begin{align}
		&\langle \bar{q}_i\nabla^\mu\nabla^\nu\nabla^\rho\nabla^\alpha\nabla^\beta q_j\rangle=\nonumber\\
		&\frac{i}{9216}\Big[-2\left(\gamma ^{\mu \nu \rho } g^{\alpha \beta }+\gamma ^{\alpha \beta \rho } g^{\mu \nu }\right) \left(2 Q_1^8-4 Q_2^8+3 Q_3^8+6 Q_5^8+4m(Q_1^7 + Q_2^7 -4 Q_3^7)+4m^2 Q^6 -4m^3 Q^5 \right)\nonumber\\[0pt]
			&-2\left(\gamma ^{\alpha \mu \nu } g^{\beta \rho }+\gamma ^{\alpha \beta \nu } g^{\mu \rho }\right) \left(2 Q_1^8+Q_3^8+2 Q_5^8+4m(Q_1^7 -2 Q_2^7-2 Q_3^7)+4m^2 Q^6-4m^3 Q^5  \right)\nonumber\\[0pt]
			&+2\left(\gamma ^{\beta \mu\nu } g^{\alpha \rho }+\gamma ^{\alpha \beta \mu } g^{\nu \rho }\right) \left(-2 Q_1^8-4 Q_2^8+Q_3^8+2 Q_5^8+4 Q_6^8-4m(Q_1^7 -2 Q_2^7)-4m^2 Q^6 +4m^3 Q^5 \right)\nonumber\\[0pt]
			&+2\gamma ^{\alpha \nu \rho } g^{\beta\mu}\left(-2 Q_1^8+2 Q_2^8+Q_3^8-2 Q_4^8+2 Q_6^8-2 m(2 Q_1^7-7 Q_2^7+2 Q_4^7)-4m^2 Q^6 +4m^3Q^5 \right)\nonumber\\[0pt]
			&+2\left(\gamma ^{\beta \nu \rho } g^{\alpha \mu }+\gamma ^{\alpha \mu \rho } g^{\beta \nu }\right) \left(\!-\!2 Q_1^8\!-\!2 Q_2^8+3 Q_3^8\!-\!2 Q_4^8+4 Q_5^8+2 Q_6^8\!-\!2 m(2 Q_1^7\!-\!Q_2^7+4 Q_3^7+2 Q_4^7)\!-\!4m^2Q^6+4m^3 Q^5\right)\nonumber\\[0pt]
			&+2\gamma ^{\beta \mu \rho } g^{\alpha \nu } \left(-2 Q_1^8+2 Q_2^8-5 Q_3^8-2 Q_4^8-12 Q_5^8+6 Q_6^8-2m(2 Q_1^7-Q_2^7+2Q_4^7-12 Q_3^7)-4 m^2Q^6 +4m^3 Q^5 \right)\nonumber\\[0pt]
			&\!+\!\left(\gamma ^{\beta }\! g^{\alpha \rho } \!g^{\mu \nu }\!\!+\!\gamma ^{\mu }\! g^{\alpha \beta }\! g^{\nu \rho }\right)\! \left(8 Q_1^8\!-\!6 Q_2^8\!+\!3 Q_3^8\!+\!2 Q_4^8\!+\!4 Q_5^8\!+\!4 Q_6^8\!+\!4m(5Q_1^7\!\!+\!3 Q_2^7\!\!+\!3 Q_3^7\!\!+\!3 Q_4^7)\!-\!20 m^2\!Q^6\!\!+\!\!12m^3\!Q^5\!\!-\!\!12 m^5\!Q^3 \right)\nonumber\\[0pt]
			&+g^{\beta \mu } \left(\gamma ^{\nu } g^{\alpha \rho }+\gamma ^{\alpha } g^{\nu \rho
			}\right) \left(8 Q_1^8+6 Q_2^8+3 Q_3^8-2 Q_4^8-8 Q_6^8-4m(7Q_1^7 -3 Q_3^7 +3 Q_4^7)+28m^2 Q^6 +12m^3 Q^5 -12m^5 Q^3 \right)\nonumber\\[0pt]
			&+\!\gamma ^{\rho }\! g^{\alpha \nu }\! g^{\beta \mu }\! \left(8 Q_1^8+2 Q_2^8+7 Q_3^8+10 Q_4^8+20 Q_5^8-20 Q_6^8-4m(7Q_1^7 - Q_2^7 -3 Q_3^7 + Q_4^7)+44m^2 Q^6 \!+\!12m^3Q^5\!-\!12m^5 Q^3 \right)\nonumber\\[0pt]
			&+\!\gamma ^{\rho }\! g^{\alpha \mu }\! g^{\beta \nu }\! \left(16 Q_1^8+2 Q_2^8-5 Q_3^8+10 Q_4^8+12 Q_5^8-20 Q_6^8+4m(Q_1^7 +Q_2^7 +3 Q_3^7 -Q_4^7)+44 m^2Q^6 \!+\!12 m^3Q^5\!-\!12m^5 Q^3 \right)\nonumber\\[0pt] 
			&-\!\left(\gamma ^{\mu }\! g^{\alpha \rho }\! g^{\beta \nu }\!\!\!+\!\!\gamma ^{\beta }\! g^{\alpha \mu }\!
			g^{\nu \rho }\right)\!\! \left(8 Q_1^8\!+\!2 Q_2^8\!+\!Q_3^8\!+\!2 Q_4^8\!+\!8 Q_5^8\!-\!8 Q_6^8\!+\!4m(11Q_1^7 \!-\!2 Q_2^7 \!-\!3 Q_3^7 \!+\!3 Q_4^7)\!-\!\!12 m^2Q^6 \!\!-\!\!12m^3 Q^5 \!\!+\!\!12m^5 Q^3 \right)\nonumber\\[0pt]
			&-\!\left(\gamma ^{\alpha }\! g^{\beta \rho }\!
			g^{\mu \nu }\!+\!\gamma ^{\nu }\! g^{\alpha \beta }\! g^{\mu \rho }\right) \!\left(8 Q_1^8\!-\!2 Q_2^8\!+\!5 Q_3^8\!-\!2 Q_4^8\!-\!4 Q_5^8\!+\!12Q_6^8\!-\!4m(Q_1^7 \!+\!Q_2^7 \!+\!3Q_3^7\!+\!3Q_4^7)\!+\!4m^2 Q^6 \!\!-\!\!12 m^3Q^5\!\!+\!\!12 m^5 Q^3\right)\nonumber\\[0pt]
			&-\gamma ^{\rho }\! g^{\alpha \beta }\! g^{\mu \nu } \!\left(24 Q_1^8\!-\!18 Q_2^8+9 Q_3^8+6 Q_4^8+12 Q_5^8+4
			Q_6^8+4m(3Q_1^7+3Q_2^7\!-\!3Q_3^7+Q_4^7)\!+\!20m^2 Q^6 \!-\!12 m^3 Q^5\!+\!12m^5 Q^3 \right)\nonumber\\[0pt]
			&+\left(\gamma ^{\nu } g^{\alpha \mu } g^{\beta \rho }+\gamma ^{\alpha } g^{\beta \nu } g^{\mu \rho }\right) \left(6 Q_2^8+3 Q_3^8\!-\!2Q_4^8+16 Q_5^8\!-\!8Q_6^8\!-\!12m(Q_1^7\!-\!Q_3^7+Q_4^7)+4(7 m^2Q^6+3 m^3Q^5\!-\!3m^5Q^3)\right)\nonumber\\[0pt]
			&-g^{\alpha \nu }\! \left(\gamma ^{\mu }\! g^{\beta \rho }+\gamma ^{\beta } \!g^{\mu \rho }\right) \!\left(2 Q_2^8\!+\!Q_3^8+2Q_4^8+8 Q_5^8\!-\!8 Q_6^8\!+\!4m(3Q_1^7\!-\!2Q_2^7\!-\!3 Q_3^7\!+\!3Q_4^7)\!-\!12(m^2 Q^6\!+\! m^3Q^5\!-\! m^5Q^3)\right)\Big]_{ji}\nonumber\\
			&-\frac{iA}{4 (d-4)(d-3)(d-2)(d-1)d}\big( \gamma^{[\mu}\gamma^\nu\gamma^\rho\gamma^\alpha\gamma^{\beta]}\big)_{ji}.\label{d8_expansion}
	\end{align}
\end{widetext}

Here, the condensate notations in Ref.~\cite{high_order_condensates} are adopted. In the last line of Eq.~\eqref{d8_expansion}, both the term $\gamma^{[\mu}\gamma^\nu\gamma^\rho\gamma^\alpha\gamma^{\beta]}$ and the denominator vanish at dimension-4; therefore, we keep this term in dimension-$d$. For convenience, the definition of each condensate is presented below. The dimension-3 and -5 condensates are simple:
\begin{equation*}
	\begin{split}
		Q^3&= \langle
		\bar{q}q\rangle, \\
		Q^5&= \langle
		\bar{q}G_{\mu\nu}\sigma^{\mu\nu}q\rangle,
	\end{split}
\end{equation*}
where $G_{\mu\nu}\equiv gT^n G^n_{\mu\nu}$ and $T^n$ is the color matrix. The dimension-6 and -7 condensates can be written as:

\begin{align}
		Q^6&=\langle\bar{q}K_\mu\gamma^\mu q\rangle= -\frac{d C_F}{4 C_A}\langle\bar{q}q\rangle ^2, \nonumber\\ 
		Q_1^7&=\langle\bar{q}G_{\mu\nu}G^{\mu\nu}q\rangle= \frac{1}{2 C_A}\langle\bar{q}q\rangle\langle GG\rangle,   \nonumber\\
		Q_2^7&=\frac{1}{2}\langle\bar{q}G_{\mu\nu}G_{\alpha\beta}\gamma^{\mu\nu\alpha\beta}q\rangle= 0,\nonumber\\
		Q_3^7&=\langle\bar{q}{G_\mu}^\lambda G_{\lambda\nu}\gamma^{\mu\nu}q\rangle= 0, \nonumber\\
		Q_4^7&=i\langle\bar{q}[\nabla_\mu,K_\nu]\gamma^{\mu\nu}q\rangle=
		-\frac{(d-1)(C_A^2-1)}{4 C_A^2}m \langle\bar{q}q\rangle ^2, \nonumber
\end{align}
where factorization (vacuum saturation hypothesis~\cite{qcd_book}) is adopted in the second equity. Similarly, the dimension-8 condensate can be factorized as:
\begin{align}
		Q_1^8&=i\langle\bar{q}[[{G_\mu}^\lambda,G_{\lambda\nu}],\nabla^\mu]\gamma^\nu q\rangle= -\frac{2}{d C_A} m\langle\bar{q}q\rangle\langle GG\rangle,  \nonumber\\
		Q_2^8&=-\frac{i}{2}\langle\bar{q}[[G_{\mu\lambda},G_{\alpha\beta}],\nabla^\mu]\gamma^{\lambda\alpha\beta} q\rangle= 0, \nonumber\\
		Q_3^8&=i\langle\bar{q}[D_\alpha G_{\mu\nu},G^{\mu\nu}]\gamma^\alpha q\rangle= -\frac{1}{2}\langle \bar{q}q\rangle  \langle
		\bar{q}Gq\rangle,\nonumber\\
		Q_5^8&=i\langle \bar{q}[G_{\mu\nu},K^\mu]\gamma^\nu q\rangle= \frac{1}{4}
		\langle \bar{q}q\rangle  \langle
		\bar{q}Gq\rangle , \nonumber\\
		Q_6^8&=\frac{i}{2}\langle\bar{q}[G_{\alpha\beta},K_\mu]\gamma^{\alpha\beta\mu}q\rangle= -\frac{(d\!-\!2)(C_A^2\!-\!2)}{8 C_A^2}\langle \bar{q}q\rangle 
		\langle \bar{q}Gq\rangle, \nonumber\\
		A&=i\langle\bar{q}\nabla_\alpha\nabla_\beta\nabla_\mu\nabla_\nu\nabla_\rho \gamma^{\alpha\beta\mu\nu\rho}q\rangle=0, \nonumber
\end{align}
where $\gamma^{\mu\nu\cdots\rho}=\gamma^{[\mu}\gamma^\nu\cdots\gamma^{\rho]}$~\cite{gamma_map}, and $K_\mu=D^\nu G_{\mu\nu}=T^n\sum_q \bar{q}\gamma_\mu T^nq$; $D^\nu$ is the covariant derivative in the adjoint representation. 

The factorization of $Q_4^8$ is ambiguous: applying the equation of motion before or after the factorization yields different results. Applying the equation of motion first yields
\begin{align}
	Q_4^8&=\langle\bar{q}[\nabla^\mu,[\nabla_\mu, K_\nu]]\gamma^\nu q\rangle\nonumber\\
	&= \frac{(d-1)
		(C_A^2\!-1)}{4 C_A^2}m^2 \langle \bar{q}q\rangle^2\!-\frac{d (C_A^2\!-2)+4}{8 C_A^2}\langle\bar{q}q\rangle\langle
		\bar{q}Gq\rangle,\nonumber
		\label{d8_em-first}
\end{align}
while applying factorization first yields
\begin{equation}
		Q_4^8= \frac{(d-1)(C_A^2-1)}{4 C_A^2}m^2 \langle\bar{q}q\rangle^2-\frac{d(C_A^2-1)}{8 C_A^2}\langle \bar{q}q\rangle\langle\bar{q}Gq\rangle.\nonumber
		\label{d8_vs-first}
\end{equation}
In this work, these two equations yield the same results because their difference vanishes at 4-dimension. 

For condensate like $\langle\bar{q}_i^a(x)q_j^b(z)\bar{q}_k^c(0)q_l^d(0)\rangle$ in
\begin{equation*}
		\begin{tikzpicture}[baseline=-\the\dimexpr\fontdimen22\textfont2\relax]
			\begin{feynman}
				\vertex (al);
				\vertex [right=2cm of al](ar);
				\vertex [right=0.7cm of al](aml);
				\vertex [right=1.3cm of al](amr);
				\vertex [right=1cm of al](am);
				\vertex [above=0.34cm of am](amu);
				\vertex [below=0.34cm of am](amd);
				\vertex [above=0.315cm of aml](amlu);
				\vertex [above=0.33cm of amr](amru);
				\vertex [below=0.315cm of amr](amrd);
				\vertex [right=0.5cm of al](amll);
				\vertex [below=0.3cm of amll](amlld);
				\vertex [right=1.5cm of al](amrr);
				\vertex [above=0.3cm of amrr](amrru);
				\vertex [right=0.18cm of al](acll);
				\vertex [below=0.17cm of acll](aclld);
				\vertex [right=0.8cm of al](acl);
				\vertex [below=0.2cm of acl](acld);
				\diagram*[small]{
					(amu)-- [ bend right=15, insertion={[size=1.5pt]0}] (al),
					(amd)--[ bend right=15, insertion={[size=1.5pt]0}](ar),
					(amrru)--[bend left=15, insertion={[size=1.5pt]0}] (ar),
					(amlld)--[bend left=15, insertion={[size=1.5pt]0}] (al),
					(amlu)--[photon](amrd),
					(acld)--[photon, bend right=22, insertion={[size=1.5pt]0}] (aclld)
				};
			\end{feynman}
		\end{tikzpicture}
		\qquad\text{and}\qquad
		\begin{tikzpicture}[baseline=-\the\dimexpr\fontdimen22\textfont2\relax]
			\begin{feynman}
				\vertex (al);
				\vertex [right=2cm of al](ar);
				\vertex [right=0.3cm of al](aml);
				\vertex [right=1.3cm of al](amr);
				\vertex [right=1cm of al](am);
				\vertex [above=0.34cm of am](amu);
				\vertex [below=0.34cm of am](amd);
				\vertex [above=0.18cm of aml](amlu);
				\vertex [above=0.33cm of amr](amru);
				\vertex [below=0.315cm of amr](amrd);
				\vertex [right=0.5cm of al](amll);
				\vertex [below=0.3cm of amll](amlld);
				\vertex [right=1.5cm of al](amrr);
				\vertex [above=0.3cm of amrr](amrru);
				\vertex [right=0.78cm of al](acl);
				\vertex [above=0.31cm of acl](aclu);
				\vertex [right=1.28cm of al](acr);
				\vertex [above=0.25cm of acr](acru);
				\diagram*[small]{
					(amu)-- [ bend right=15, insertion={[size=1.5pt]0}] (al),
					(amd)--[ bend right=15, insertion={[size=1.5pt]0}](ar),
					(amrru)--[bend left=15, insertion={[size=1.5pt]0}] (ar),
					(amlld)--[bend left=15, insertion={[size=1.5pt]0}] (al),
					(amlu)--[photon](amrd),
					(acru)--[photon, bend left=25, insertion={[size=1.5pt]0}] (aclu)
				};
			\end{feynman}
		\end{tikzpicture},
\end{equation*}
we factorize it before applying the equation of motion for simplicity, i.e.
\setlength{\jot}{0pt}
\begin{equation*}
	\begin{split}
	\langle&\bar{q}_i^a(x)q_j^b(z)\bar{q}_k^c(0)q_l^d(0)\rangle\\
	&=\frac{1}{2^4C_A^2}\big(\delta^{ab}_{ij}\delta^{cd}_{kl}\langle\bar{q}(x)q(z)\rangle\langle\bar{q}(0)q(0)\rangle\\
	&\qquad\qquad\qquad - \delta^{ad}_{il}\delta^{cb}_{kj}\langle\bar{q}(x)q(0)\rangle\langle\bar{q}(0)q(z)\rangle\big)\\
	&=\frac{1}{2^6 dC_A^2}\big((x-z)^2\delta^{ab}_{ij}\delta^{cd}_{kl}\langle\bar{q}Gq\rangle\langle\bar{q}q\rangle\\
	&\qquad\qquad\qquad - (x^2+z^2)\delta^{ad}_{il}\delta^{cb}_{kj} \langle\bar{q}Gq\rangle\langle\bar{q}q\rangle\big).
	\end{split}	
\end{equation*}
Applying the equation of motion before factorization yields results with difference $\propto O(1/C_A^2)$~\cite{va_sum}. For the results in Tables.~\ref{GSR_1-pole} and \ref{GSR_2-pole}, the dimension-8 condensate contribution is negligible (e.g., the lower plot in Fig.~\ref{GSR_OPE}). Therefore, the results in this work are not affected by how to factorize the dimension-8 condensate.

\setlength{\jot}{3pt}
\section{Renormalization of $\overline{\Psi}\Gamma\overleftrightarrow{\nabla}\!_\alpha\Psi$ Operator\label{ren_}}
Here we explain how Eq.~\eqref{j_ren} is obtained. At one-loop level, one can derive renormalized operator $\overline{\Psi}\Gamma\overleftrightarrow{\nabla}\!_\alpha\Psi$ for generic gamma-matrices. Consider
\begin{equation}
	\overline{\Psi}\Gamma\overleftrightarrow{\nabla}\!\!_\alpha\Psi=\overline{\Psi}\Gamma\overleftrightarrow{\partial}\!_\alpha\Psi-2ig\overline{\Psi}\Gamma A_\alpha\Psi,
	\label{operator_expand}
\end{equation}
where $A_\alpha \equiv \text{T}^nA^n_\alpha$, and all fields are renormalized implicitly. Note that the first diagram in Fig.~\ref{ren_d0} is related only to the $A_\alpha\Psi$ part of Eq.~\eqref{operator_expand}, as
\begin{equation}
	\begin{tikzpicture}[baseline=(current bounding box.center)]
		\begin{feynman}
			\vertex (a0);
			\coordinate [below=1.15cm of a0](a00);
			\vertex [left=0.9cm of a00](aLL);
			\vertex [right=0.9cm of a00](aRR);
			\coordinate [below=0.75cm of a0](au);
			\vertex [right=0.6cm of au](aur);
			\diagram*[small]{
				(aRR)--[fermion](a0)--[fermion](aLL),
				(aur)--[photon,bend left=55](a0),
			};
			\filldraw(aur) circle(0.03cm);
		\end{feynman}
	\end{tikzpicture}\sim\overline{\Psi}\Gamma\wick[offset=1.1em]{\c1{A}_\alpha \c2{\Psi} \c2{\overline{\Psi}}\c1{\slashed{A}}\Psi}.
\end{equation}
Therefore, the corresponding counterterm can be written as
\begin{equation}
	\frac{1}{\varepsilon}\overline{\Psi}\Gamma\times\big\{\text{operators act on}\, \Psi\big\}.
\end{equation}
Similarly for the second diagram in Fig.~\ref{ren_d0}, which relates only to the $\overline{\Psi}A_\alpha$ part of Eq.~\eqref{operator_expand} ($\text{T}^n$ commutes with $\Gamma$). In both cases, the counterterms for generic $\Gamma$ can be easily obtained. These two diagrams give the second and third lines in Eq.~\eqref{j_ren}.

The third diagram in Fig.~\ref{ren_d0} is more complicated:
\begin{equation}
	\begin{tikzpicture}[baseline=(current bounding box.center)]
		\begin{feynman}
			\vertex (a0);
			\coordinate [below=1.15cm of a0](a00);
			\vertex [left=0.9cm of a00](aLL);
			\vertex [right=0.9cm of a00](aRR);
			\coordinate [below=0.75cm of a0](au);
			\vertex [left=0.6cm of au](aul);
			\vertex [right=0.6cm of au](aur);
			\diagram*[small]{
				(aRR)--[fermion](a0)--[fermion](aLL),
				(aul)--[photon](aur),
			};
			\draw(0,0.24) node[scale=0.75, transform shape, rotate=90]{$\leftarrow$};
			\draw(0.4,0.22) node[scale=0.75, transform shape]{$p-q$};
			\draw(0.64,-1.2) node[scale=0.75, transform shape, rotate=-54]{$\leftarrow q$};
			\draw(-0.64,-1.2) node[scale=0.75, transform shape, rotate=54]{$\leftarrow p$};
			\filldraw(aur) circle(0.03cm);
			\filldraw(aul) circle(0.03cm);
		\end{feynman}
	\end{tikzpicture}\sim\wick[offset=1.1em]{\overline{\Psi}\c1{\slashed{A}}\c2{\Psi}\big(\c2{\overline{\Psi}}\Gamma\overleftrightarrow{\partial}\!_\alpha \c2{\Psi}\big)\c2{\overline{\Psi}}\c1{\slashed{A}}\Psi},
	\label{ren_psi-A-psi}
\end{equation}
where $p\neq q$ for generic. Leaving $\Gamma$ unspecified, Eq.~\eqref{ren_psi-A-psi} then yields:
\begin{align}
	&\wick[offset=1.1em]{\c1{\slashed{A}}\c2{\Psi}_{f_a}\big(\c2{\overline{\Psi}}_{f_a}\Gamma\overleftrightarrow{\partial}\!_\alpha \c2{\Psi}_{f_b}\big)\c2{\overline{\Psi}}_{f_b}\c1{\slashed{A}}}=\nonumber\\
	&\quad\quad\frac{iC_F}{192\varepsilon\pi^2}(p_\alpha + q_\alpha)\, \gamma^\mu\gamma^\nu\Gamma\gamma_\nu\gamma_\mu\nonumber\\
	&\quad\quad+\frac{iC_F}{96\varepsilon\pi^2}\Big(\gamma^\mu(2\slashed{p}-\slashed{q})\Gamma\gamma_\alpha\gamma_\mu + \gamma^\mu\gamma_\alpha\Gamma(2\slashed{q}-\slashed{p})\gamma_\mu\Big)\nonumber\\
	&\quad\quad+\frac{iC_F}{32\varepsilon\pi^2}\big(m_{f_a}\gamma^\mu\Gamma\gamma_\alpha\gamma_\mu + \gamma^\mu\gamma_\alpha\Gamma\gamma_\mu\, m_{f_b}\big)\nonumber\\
	&\quad\quad+O(1),
\end{align}
which gives the last four lines in Eq.~\eqref{j_ren}. This technique has been used in Refs.~\cite{li2025qcdsumruleanalysis, Li2026}. Writing the $\Gamma$ explicitly then yields the renormalized current for each current in Eq.~\eqref{current_basis}. 

\section{Monte Carlo Uncertainty Analysis\label{_montecarlo}}

It should be noted that some QCD parameters involve large uncertainties (e.g., $\langle GG\rangle = 0.07\pm0.02\,\text{GeV}^4$, $M_0^2=0.8\pm0.2\,\text{GeV}^2$); thus, the actual errors of the results cannot be reliably obtained by the standard linear propagation of uncertainty. Therefore, a Monte Carlo uncertainty analysis is adopted to estimate the errors. The procedure adopted here is the same as that in Ref.~\cite{Li2026}; below, we detail how it is performed for the two-resonance GSR analysis.

Based on the QCD parameters listed in Section.~\ref{currents_basis} (where $\langle \bar{q}q\rangle$ and $\langle \bar{s}s\rangle$ are assigned a 10\% relative uncertainty) and the masses given in the PDG~\cite{pdg}, we generate 2000 sets of input parameters by sampling from independent Gaussian distributions. For each parameter set, we obtain $m$ ($m_1$ or $m_2$), $r$, and $s_0$ by minimizing Eq.~\eqref{gsr_chi} with $\widetilde{G}_h(\tau,s)$ replaced by Eq.~\eqref{2-deltas}. We then obtain 2000 optimal sets of $(m, r, s_0)$. The uncertainties are combined from two sources:
\begin{itemize}
	\item $\hat{\sigma}^2_m$, $\hat{\sigma}^2_r$, and $\hat{\sigma}^2_{s_0}$: The variance of the 2000 fitted values, which originate from the uncertainties of the QCD parameters.
	
	\item $\overline{\Delta^2_m}$, $\overline{\Delta^2_r}$, and $\overline{\Delta^2_{s_0}}$: The mean of the squared fitting errors from each individual fit. The fitting errors $\Delta_m$, $\Delta_r$, and $\Delta_{s_0}$ for a single fit are estimated from the Hessian matrix
	\NiceMatrixOptions{cell-space-limits = 0.05cm}
	\begin{equation}
		H(\chi^2)=\begin{bmatrix}
			\displaystyle\frac{\partial^2}{\partial m^2}& \displaystyle\frac{\partial^2}{\partial m\partial r} & \displaystyle\frac{\partial^2}{\partial m\partial s_0}\\
			\displaystyle\frac{\partial^2}{\partial r\partial m}&\displaystyle\frac{\partial^2}{\partial r^2}& \displaystyle\frac{\partial^2}{\partial r\partial s_0}\\
			\displaystyle\frac{\partial^2}{\partial s_0\partial m}& \displaystyle\frac{\partial^2}{\partial s_0\partial r}& \displaystyle\frac{\partial^2}{\partial s^2_0}\\
		\end{bmatrix} \chi^2
	\end{equation}
	evaluated at the minimum of $\chi^2$, via
	\begin{equation}
		\Delta=\sqrt{2\chi^2_\text{min}\, H^{-1}(\chi^2_\text{min})}.
		\label{delta_err}
	\end{equation}
	The diagonal components of $\Delta$ then give $\Delta_m$, $\Delta_r$, and $\Delta_{s_0}$. 
\end{itemize}
The value of Eq.~\eqref{delta_err} is independent of the $\delta s$ in Eq.~\eqref{gsr_chi}, provided that $\delta s$ is chosen sufficiently small. In the limit $\delta s\to0$, $\chi^2$ scales as $1/\delta s$, while $H^{-1}$ scales as $\delta s$. It is also independent of the choices of $s_{\text{min}}$ and $s_{\text{max}}$, provided that $s_{\text{min}}$ and $s_{\text{max}}$ are far away from the peak, where both $\widetilde{G}(\tau,s,s_0)$ and $\widetilde{G}_h(\tau,s)$ are essentially zero. Eq.~\eqref{delta_err} quantifies the intrinsic uncertainty of the fitting method itself.

The total uncertainties $\sigma_m, \sigma_r, \sigma_{s_0}$ are then computed by combining the two distinct sources of error in quadrature:
\begin{equation}
	\sigma=\sqrt{\hat{\sigma}^2 + \overline{\Delta^2}}.
\end{equation}

\section{OPE Convergence, and Sensitivity to the Choice of $\tau$ in GSR\label{ope_tau}}

\begin{figure}[t!]
	\centering
	\includegraphics[width=6.5cm, height=4cm]{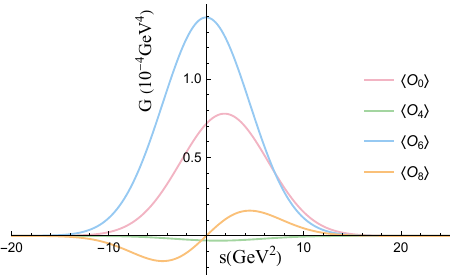}\\
	\vspace{0.36cm}
	\includegraphics[width=6.5cm, height=4cm]{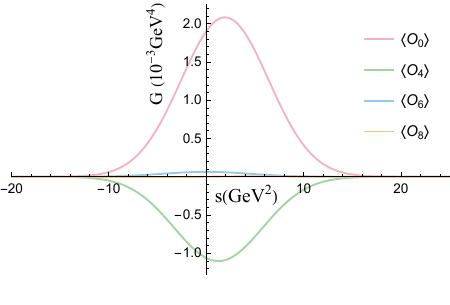}\\
	\vspace{-0.2cm}
	\caption{Condensates' contributions after Gaussian transformation; $\langle O_n\rangle$ denotes combined contributions for dimension-$n$ condensates; $s_0=2.5\text{GeV}^2$ is used. {\bf{Upper:}} $\bar{u}d$ state in the $J^{PC}=1^{++}$ channel corresponding to $J^{\mu\nu\alpha}$ ($\epsilon^{\mu\nu\rho\sigma}\epsilon_\rho p_\sigma p^\alpha$ polarization tensor). The inclusion of the dimension-8 condensate is necessary to ensure the OPE truncation is valid; the $\langle O_4\rangle$ contribution is nearly negligible because $\langle GG\rangle$ contribution is absent; the condensates' contribution is dominant due to the absence of LO perturbative contribution. {\bf{Lower:}} $\bar{u}d$ state in the $J^{PC}=2^{--}$ channel corresponding to $\widetilde{J}^{\{\mu\alpha\}}$; the OPE is well convergent; dimension-8 condensate contribution is negligible.}
	\label{GSR_OPE}
\end{figure}

\begin{figure}[h!]
	\centering
	\includegraphics[width=6.85cm, height=4.5cm]{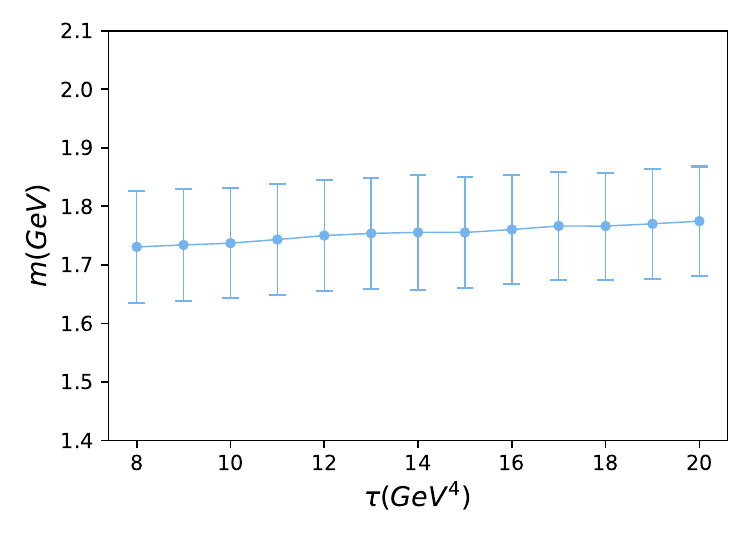}
	\vspace{-0.4cm}
	\caption{GSR mass prediction versus $\tau (\text{GeV}^4)$ for the $\bar{u}d$ state with $J^{PC}=2^{--}$, corresponding to $\widetilde{J}^{\{\mu\alpha\}}$.}
	\label{GSR_tau}
\end{figure}

As shown in Fig.~\ref{GSR_OPE}, the inclusion of dimension-8 condensates ensures the truncation of OPE is valid; Fig.~\ref{GSR_tau} shows that the predicted masses are not sensitive to the choice of $\tau$.

\section{Correlators\label{corr_results}}

Here we present the OPE results for each correlators. In each case, the Lorentz projector (using Eqs.~\eqref{polarization-2-3} and \eqref{levi_civita_sum}) is normalized and multiplied by the factor $p^{-2n}$ to make it dimensionless; the overall sign is determined by requiring $\text{Im}\Pi(s)>0$ as $s\rightarrow\infty$. Contributions from condensates with dimension $\leq4$ are presented in Table~\ref{corre_d4}, while condensates with dimension $\geq 6$ are presented in Table~\ref{corre_d6-d8}. Results for the $\bar{u}s$ configuration are obtained by replacing $d\rightarrow s$; replacing $u,d\rightarrow s$ and combining the same terms yields the results for the $\bar{s}s$ configuration. The leading-order perturbative contributions are evaluated using massive quark propagators, expanding the results to $O(m^2)$.

\break

\begin{turnpage}
	\begin{table*}
		\caption{Correlators for each current in the $\bar{u}d$ configuration; $t=\log (-\frac{p^2}{\mu^2})$ here. The second column lists the polarization tensors for the corresponding quantum numbers, where square and curly brackets denote anti-symmetrization and symmetrization of the indices, respectively.}
		\label{corre_d4}
		\renewcommand{\arraystretch}{1.4}
		\begin{ruledtabular}
		\begin{tabular}{clccccccccccccc}
			&&\multirow{2}{*}{$\frac{p^4 t}{\pi^2}$}&\multirow{2}{*}{$\frac{(m_u^2\!+m_d^2)p^2 t}{\pi^2}$}&\multirow{2}{*}{$\frac{m_um_dp^2 t}{\pi^2}$}&\multirow{2}{*}{$\frac{\alpha_sp^4 t}{\pi^3}$}&\multirow{2}{*}{$\frac{\alpha_sp^4 t^2}{\pi^3}$}&$\frac{m_u\langle\bar{d}d\rangle\alpha_st}{\pi}$&$\frac{m_u\langle\bar{u}u\rangle\alpha_st}{\pi}$&$\frac{\alpha_sm_u\langle\bar{d}d\rangle}{\pi}$&$\frac{\alpha_sm_u\langle\bar{u}u\rangle}{\pi}$&$\frac{m_u\langle\bar{d}d\rangle\,\,}{\pi}$&$\frac{m_u\langle\bar{u}u\rangle\,\,}{\pi}$&\multirow{2}{*}{$\frac{\langle GG\rangle t}{\pi}$}&\multirow{2}{*}{$\frac{\langle GG\rangle}{\pi}$}\\[-4pt]
			&&&&&&&{\tiny$+\{u\leftrightarrow d\}$}&{\tiny$+\{u\leftrightarrow d\}$}&{\tiny$+\{u\leftrightarrow d\}$}&{\tiny$+\{u\leftrightarrow d\}$}&{\tiny$+\{u\leftrightarrow d\}$}&{\tiny$+\{u\leftrightarrow d\}$}&&\\
			\hline
			\multirow{2}{*}{$J^\mu$}&$0^{+-}:p^\mu$&$0$&$0$&$0$&$0$&$0$&$-\frac{1}{3}$&$-\frac{1}{6}$&$\frac{2}{3}$&$\frac{1}{4}$&$-1$&$-\frac{3}{2}$&$0$&$0$\\
			&$1^{--}:\epsilon^\mu$&$-\frac{1}{8}$&$\frac{1}{2}$&$\frac{1}{4}$&$-\frac{11}{72}$&$-\frac{1}{24}$&$-\frac{5}{3}$&$-\frac{13}{6}$&$\frac{8}{3}$&$\frac{7}{4}$&$0$&$0$&$0$&$-\frac{1}{24}$\\ \hline
			\multirow{2}{*}{$\widetilde{J}^\mu$}&$0^{--}:p^\mu$&$0$&$0$&$0$&$0$&$0$&$-\frac{1}{3}$&$\frac{1}{6}$&$\frac{2}{3}$&$-\frac{1}{4}$&$-1$&$\frac{3}{2}$&$0$&$0$\\ 
			&$1^{+-}:\epsilon^\mu$&$-\frac{1}{8}$&$\frac{1}{2}$&$-\frac{1}{4}$&$-\frac{11}{72}$&$-\frac{1}{24}$&$\frac{5}{3}$&$-\frac{13}{6}$&$-\frac{8}{3}$&$\frac{7}{4}$&$0$&$0$&$0$&$-\frac{1}{24}$\\ \hline
			\multirow{4}{*}{$J^{\{\mu\alpha\}}$}&$0^{++}:\eta^{\mu\alpha}$&$0$&$-\frac{1}{24}$&$-\frac{1}{12}$&$-\frac{1}{1080}$&$0$&$\frac{1}{9}$&$\frac{19}{108}$&$0$&$-\frac{577}{648}$&$0$&$0$&$\frac{1}{54}$&$-\frac{5}{324}$\\
			&$0^{++}:p^\mu p^\alpha$&$0$&$0$&$0$&$-\frac{1}{120}$&$0$&$\frac{1}{3}$&$\frac{1}{12}$&$0$&$-\frac{13}{24}$&$-1$&$0$&$\frac{1}{6}$&$-\frac{13}{36}$\\
			&$1^{-+}:\epsilon^{\{\mu} p^{\alpha\}}$&$0$&$-\frac{1}{32}$&$\frac{1}{16}$&$-\frac{1}{60}$&$0$&$-\frac{13}{18}$&$\frac{29}{72}$&$\frac{31}{27}$&$-\frac{389}{432}$&$\frac{1}{4}$&$0$&$-\frac{1}{9}$&$\frac{11}{54}$\\ 
			&$2^{++}:\epsilon^{\mu\alpha}$&$-\frac{3}{40}$&$\frac{1}{4}$&$-\frac{1}{4}$&$\frac{413}{1800}$&$-\frac{1}{15}$&$\frac{7}{3}$&$-\frac{29}{36}$&$-4$&$-\frac{61}{216}$&$0$&$0$&$\frac{2}{9}$&$-\frac{49}{216}$\\ \hline
			\multirow{2}{*}{$J^{[\mu\alpha]}$}&$1^{-+}:\epsilon^{[\mu} p^{\alpha]}$&$0$&$-\frac{1}{32}$&$\frac{1}{16}$&$0$&$0$&$\frac{1}{6}$&$-\frac{1}{24}$&$-\frac{1}{3}$&$\frac{1}{16}$&$\frac{1}{4}$&$-\frac{1}{2}$&$0$&$0$\\
			&$1^{++}:\epsilon^{\mu\alpha\rho\sigma} \epsilon_\rho p_\sigma$&$-\frac{1}{16}$&$\frac{1}{4}$&$-\frac{1}{8}$&$-\frac{29}{144}$&$0$&$\frac{1}{2}$&$-\frac{5}{8}$&$-\frac{2}{3}$&$-\frac{13}{48}$&$0$&$0$&$0$&$\frac{1}{48}$\\ \hline 
			\multirow{4}{*}{$\widetilde{J}^{\{\mu\alpha\}}$}&$0^{--}:\eta^{\mu\alpha}$&$0$&$-\frac{1}{24}$&$\frac{1}{12}$&$-\frac{1}{1080}$&$0$&$-\frac{1}{9}$&$\frac{19}{108}$&$0$&$-\frac{577}{648}$&$0$&$0$&$\frac{1}{54}$&$-\frac{5}{324}$\\
			&$0^{--}:p^\mu p^\alpha$&$0$&$0$&$0$&$-\frac{1}{120}$&$0$&$-\frac{1}{3}$&$\frac{1}{12}$&$0$&$-\frac{13}{24}$&$1$&$0$&$\frac{1}{6}$&$-\frac{13}{36}$\\
			&$1^{+-}:\epsilon^{\{\mu} p^{\alpha\}}$&$0$&$-\frac{1}{32}$&$-\frac{1}{16}$&$-\frac{1}{60}$&$0$&$\frac{13}{18}$&$\frac{29}{72}$&$-\frac{31}{27}$&$-\frac{389}{432}$&$-\frac{1}{4}$&$0$&$-\frac{1}{9}$&$\frac{11}{54}$\\ 
			&$2^{--}:\epsilon^{\mu\alpha}$&$-\frac{3}{40}$&$\frac{1}{4}$&$\frac{1}{4}$&$\frac{413}{1800}$&$-\frac{1}{15}$&$-\frac{7}{3}$&$-\frac{29}{36}$&$4$&$-\frac{61}{216}$&$0$&$0$&$\frac{2}{9}$&$-\frac{49}{216}$\\ \hline
			\multirow{2}{*}{$\widetilde{J}^{[\mu\alpha]}$}&$1^{+-}:\epsilon^{[\mu} p^{\alpha]}$&$0$&$-\frac{1}{32}$&$-\frac{1}{16}$&$0$&$0$&$-\frac{1}{6}$&$-\frac{1}{24}$&$\frac{1}{3}$&$\frac{1}{16}$&$-\frac{1}{4}$&$-\frac{1}{2}$&$0$&$0$\\
			&$1^{--}:\epsilon^{\mu\alpha\rho\sigma} \epsilon_\rho p_\sigma$&$-\frac{1}{16}$&$\frac{1}{4}$&$\frac{1}{8}$&$-\frac{29}{144}$&$0$&$-\frac{1}{2}$&$-\frac{5}{8}$&$\frac{2}{3}$&$-\frac{13}{48}$&$0$&$0$&$0$&$\frac{1}{48}$\\ \hline
			\multirow{8}{*}{$J^{\mu\nu\alpha}$}&$0^{-+}:\epsilon^{\mu\nu\alpha\rho}p_\rho$&$-\frac{1}{24}$&$\frac{1}{6}$&$\frac{1}{12}$&$-\frac{29}{72}$&$\frac{1}{24}$&$-\frac{1}{9}$&$\frac{1}{9}$&$0$&$-\frac{3}{2}$&$0$&$0$&$0$&$-\frac{1}{72}$\\
			&$0^{++}:\eta^{\mu\alpha}p^\nu - \eta^{\nu\alpha}p^\mu$&$-\frac{1}{24}$&$\frac{1}{6}$&$-\frac{1}{12}$&$-\frac{29}{72}$&$\frac{1}{24}$&$\frac{1}{9}$&$\frac{1}{9}$&$0$&$-\frac{8}{27}$&$0$&$0$&$0$&$-\frac{1}{72}$\\
			&$1^{-+}:\eta^{\mu\alpha}\epsilon^\nu - \eta^{\nu\alpha}\epsilon^\mu$&$0$&$-\frac{1}{16}$&$\frac{1}{8}$&$-\frac{1}{240}$&$0$&$-\frac{1}{2}$&$\frac{5}{9}$&$\frac{2}{3}$&$-\frac{211}{108}$&$0$&$0$&$0$&$\frac{1}{36}$\\
			&$1^{-+}:\epsilon^\mu p^\nu p^\alpha\!\!-\!\epsilon^\nu p^\mu p^\alpha$&$0$&$0$&$0$&$-\frac{1}{60}$&$0$&$\frac{1}{3}$&$-\frac{4}{9}$&$-\frac{10}{9}$&$\frac{20}{27}$&$1$&$-\frac{1}{2}$&$0$&$\frac{1}{9}$\\
			&$1^{++}:\epsilon^{\mu\nu\rho\sigma} \epsilon_\rho p_\sigma p^\alpha$&$0$&$0$&$0$&$-\frac{1}{60}$&$0$&$-\frac{1}{3}$&$-\frac{4}{9}$&$\frac{2}{9}$&$-\frac{17}{54}$&$-1$&$-\frac{1}{2}$&$0$&$-\frac{2}{9}$\\ [2pt]
			&$1^{++}:{\epsilon^{\mu\alpha\rho\sigma} \epsilon_\rho p_\sigma p^\nu \atop -\{\mu\leftrightarrow\nu\}}$&$0$&$-\frac{1}{16}$&$-\frac{1}{8}$&$-\frac{1}{240}$&$0$&$\frac{1}{2}$&$\frac{5}{9}$&$-\frac{2}{3}$&$-\frac{25}{27}$&$0$&$0$&$0$&$-\frac{1}{18}$\\[2pt]
			&$2^{++}:\epsilon^{\mu\alpha}p^\nu - \epsilon^{\nu\alpha}p^\mu$&$-\frac{1}{20}$&$\frac{1}{8}$&$-\frac{1}{4}$&$\frac{107}{600}$&$-\frac{1}{20}$&$\frac{7}{3}$&$0$&$-4$&$0$&$0$&$0$&$0$&$-\frac{1}{12}$\\
			&$2^{-+}:\epsilon^{\mu\nu\rho\sigma}{\epsilon_\rho}^\alpha p_\sigma$&$-\frac{1}{20}$&$\frac{1}{8}$&$\frac{1}{4}$&$\frac{107}{600}$&$-\frac{1}{20}$&$-\frac{7}{3}$&$0$&$4$&$-\frac{29}{18}$&$0$&$0$&$0$&$\frac{5}{12}$
		\end{tabular}
		\end{ruledtabular}
	\end{table*}
\end{turnpage}


\begin{turnpage}
	\begin{table*}
		\caption{Correlators for each current in the $\bar{u}d$ configuration, for condensates with dimension $\geq6$; the same conventions as in Table~\ref{corre_d4} are adopted. The last two columns vanish for the $\bar{f_a}f_b$ configuration with $f_a\neq f_b$.}
		\label{corre_d6-d8}
		\renewcommand{\arraystretch}{1.4}
		\begin{ruledtabular}
			\begin{tabular}{clccccccccccc}
				&&$\frac{m_u\langle\bar{d}Gd\rangle}{p^2}$&$\frac{m_u\langle\bar{u}Gu\rangle}{p^2}$&$\frac{\pi\alpha_s\langle\bar{u}u\rangle^2}{p^2}$&\multirow{2}{*}{$\frac{\pi\alpha_s\langle\bar{u}u\rangle\langle\bar{d}d\rangle}{p^2}$}&\multirow{2}{*}{$\frac{\langle GGG\rangle}{\pi^2p^2}$}&$\frac{\pi m_u\langle\bar{d}d\rangle\langle GG\rangle}{p^4}$&$\frac{\pi m_u\langle\bar{u}u\rangle\langle GG\rangle}{p^4}$&$\frac{\pi\alpha_s \langle\bar{u}u\rangle\langle\bar{u}Gu\rangle}{p^4}$&$\frac{\pi\alpha_s \langle\bar{u}u\rangle\langle\bar{d}Gd\rangle}{p^4}$&$\frac{\pi\alpha_s \langle\bar{u}u\rangle\langle\bar{u}Gu\rangle}{p^4}$&$\frac{\pi\alpha_s \langle\bar{d}d\rangle\langle\bar{d}Gd\rangle}{p^4}$\\[-4pt]
				&&{\tiny$+\{u\leftrightarrow d\}$}&{\tiny$+\{u\leftrightarrow d\}$}&{\tiny$+\{u\leftrightarrow d\}$}&&&{\tiny$+\{u\leftrightarrow d\}$}&{\tiny$+\{u\leftrightarrow d\}$}&{\tiny$+\{u\leftrightarrow d\}$}&{\tiny$+\{u\leftrightarrow d\}$}&{\tiny$\times\delta_{ud}$}&{\tiny$\times\delta_{ud}$}\\
				\hline
				\multirow{2}{*}{$J^\mu$}&$0^{+-}:p^\mu$&$0$&$0$&$\frac{2}{27}$&$0$&$0$&$-\frac{5}{6}$&$-\frac{1}{12}$&$\frac{71}{72}$&$\frac{31}{1620}$&$0$&$0$\\
				&$1^{--}:\epsilon^\mu$&$-\frac{1}{2}$&$-\frac{1}{3}$&$\frac{10}{27}$&$\frac{16}{9}$&$\frac{1}{144}$&$-\frac{5}{18}$&$-\frac{1}{36}$&$-\frac{91}{216}$&$\frac{203}{2430}$&$0$&$0$\\ \hline
				\multirow{2}{*}{$\widetilde{J}^\mu$}&$0^{--}:p^\mu$&$0$&$0$&$-\frac{2}{27}$&$0$&$0$&$-\frac{5}{6}$&$\frac{1}{12}$&$\frac{71}{72}$&$-\frac{31}{1620}$&$0$&$0$\\ 
				&$1^{+-}:\epsilon^\mu$&$\frac{1}{2}$&$-\frac{1}{3}$&$\frac{10}{27}$&$-\frac{16}{9}$&$\frac{1}{144}$&$\frac{5}{18}$&$-\frac{1}{36}$&$\frac{91}{216}$&$\frac{203}{2430}$&$0$&$0$\\ \hline
				\multirow{4}{*}{$J^{\{\mu\alpha\}}$}&$0^{++}:\eta^{\mu\alpha}$&$\frac{5}{36}$&$\frac{1}{6}$&$\frac{2}{27}$&$-\frac{16}{27}$&$-\frac{1}{432}$&$\frac{13}{162}$&$\frac{2}{27}$&$-\frac{67}{648}$&$-\frac{613}{4860}$&$0$&$0$\\
				&$0^{++}:p^\mu p^\alpha$&$-\frac{1}{2}$&$\frac{1}{2}$&$\frac{2}{3}$&$-\frac{16}{3}$&$-\frac{1}{48}$&$-\frac{19}{18}$&$\frac{1}{3}$&$-\frac{133}{72}$&$-\frac{791}{810}$&$0$&$0$\\
				&$1^{-+}:\epsilon^{\{\mu} p^{\alpha\}}$&$\frac{1}{24}$&$\frac{1}{48}$&$-\frac{68}{81}$&$\frac{16}{9}$&$\frac{1}{48}$&$\frac{1}{3}$&$\frac{5}{27}$&$\frac{25}{216}$&$-\frac{55}{96}$&$0$&$0$\\ 
				&$2^{++}:\epsilon^{\mu\alpha}$&$\frac{13}{24}$&$0$&$0$&$-\frac{16}{9}$&$-\frac{1}{36}$&$\frac{8}{27}$&$\frac{1}{9}$&$\frac{23}{54}$&$-\frac{1223}{6480}$&$0$&$0$\\ \hline
				\multirow{2}{*}{$J^{[\mu\alpha]}$}&$1^{-+}:\epsilon^{[\mu} p^{\alpha]}$&$-\frac{1}{24}$&$-\frac{1}{48}$&$-\frac{2}{81}$&$0$&$0$&$\frac{7}{54}$&$-\frac{1}{27}$&$-\frac{43}{216}$&$-\frac{571}{38880}$&$0$&$0$\\
				&$1^{++}:\epsilon^{\mu\alpha\rho\sigma} \epsilon_\rho p_\sigma$&$\frac{11}{48}$&$-\frac{1}{6}$&$-\frac{2}{27}$&$-\frac{8}{9}$&$0$&$\frac{7}{54}$&$-\frac{1}{27}$&$\frac{31}{72}$&$\frac{89}{38880}$&$0$&$0$\\ \hline 
				\multirow{4}{*}{$\widetilde{J}^{\{\mu\alpha\}}$}&$0^{--}:\eta^{\mu\alpha}$&$-\frac{5}{36}$&$\frac{1}{6}$&$\frac{2}{27}$&$\frac{16}{27}$&$-\frac{\
					1}{432}$&$-\frac{13}{162}$&$\frac{2}{27}$&$\frac{67}{648}$&$-\frac{\
					613}{4860}$&$0$&$0$\\
				&$0^{--}:p^\mu p^\alpha$&$\frac{1}{2}$&$\frac{1}{2}$&$\frac{2}{3}$&$\frac{16}{3}$&$-\frac{1}{48}$&$\frac{19}{18}$&$\frac{1}{3}$&$\frac{133}{72}$&$-\frac{791}{810}$&$0$&$0$\\
				&$1^{+-}:\epsilon^{\{\mu} p^{\alpha\}}$&$-\frac{1}{24}$&$\frac{1}{48}$&$-\frac{68}{81}$&$-\frac{16}{9}$&$\frac{1}{48}$&$-\frac{1}{3}$&$\frac{5}{27}$&$-\frac{25}{216}$&$-\frac{55}{96}$&$0$&$0$\\ 
				&$2^{--}:\epsilon^{\mu\alpha}$&$-\frac{13}{24}$&$0$&$0$&$\frac{16}{9}$&$-\frac{1}{36}$&$-\frac{8}{27}$&$\frac{1}{9}$&$-\frac{23}{54}$&$-\frac{1223}{6480}$&$0$&$0$\\ \hline
				\multirow{2}{*}{$\widetilde{J}^{[\mu\alpha]}$}&$1^{+-}:\epsilon^{[\mu} p^{\alpha]}$&$\frac{1}{24}$&$-\frac{1}{48}$&$-\frac{2}{81}$&$0$&$0$&$-\frac{7}{54}$&$-\frac{1}{27}$&$\frac{43}{216}$&$-\frac{571}{38880}$&$0$&$0$\\
				&$1^{--}:\epsilon^{\mu\alpha\rho\sigma} \epsilon_\rho p_\sigma$&$-\frac{11}{48}$&$-\frac{1}{6}$&$-\frac{2}{27}$&$\frac{8}{9}$&$0$&$-\frac{7}{54}$&$-\frac{1}{27}$&$-\frac{31}{72}$&$\frac{89}{38880}$&$0$&$0$\\ \hline
				\multirow{8}{*}{$J^{\mu\nu\alpha}$}&$0^{-+}:\epsilon^{\mu\nu\alpha\rho}p_\rho$&$-\frac{5}{36}$&$-\frac{1}{36}$&$-\frac{2}{27}$&$\frac{16}{27}$&$0$&$-\frac{13}{162}$&$-\frac{11}{324}$&$-\frac{103}{216}$&$-\frac{163}{29160}$&$\frac{4}{27}$&$\frac{2}{27}$\\
				&$0^{++}:\eta^{\mu\alpha}p^\nu - \eta^{\nu\alpha}p^\mu$&$\frac{5}{36}$&$-\frac{1}{36}$&$-\frac{2}{27}$&$-\frac{16}{27}$&$0$&$\frac{13}{162}$&$-\frac{11}{324}$&$\frac{103}{216}$&$-\frac{163}{29160}$&$-\frac{2}{27}$&$-\frac{1}{27}$\\
				&$1^{-+}:\eta^{\mu\alpha}\epsilon^\nu - \eta^{\nu\alpha}\epsilon^\mu$&$-\frac{11}{48}$&$\frac{5}{16}$&$-\frac{1}{27}$&$\frac{8}{9}$&$0$&$-\frac{7}{54}$&$\frac{25}{108}$&$\frac{1}{3}$&$-\frac{8141}{38880}$&$-\frac{1}{9}$&$-\frac{1}{18}$\\
				&$1^{-+}:\epsilon^\mu p^\nu p^\alpha\!\!-\!\epsilon^\nu p^\mu p^\alpha$&$\frac{2}{3}$&$0$&$-\frac{10}{81}$&$\frac{32}{9}$&$0$&$\frac{23}{18}$&$\frac{7}{36}$&$\frac{239}{216}$&$-\frac{4183}{4860}$&$-\frac{1}{9}$&$-\frac{4}{9}$\\
				&$1^{++}:\epsilon^{\mu\nu\rho\sigma} \epsilon_\rho p_\sigma p^\alpha$&$-\frac{2}{3}$&$0$&$-\frac{10}{81}$&$-\frac{32}{9}$&$0$&$-\frac{23}{18}$&$\frac{7}{36}$&$-\frac{239}{216}$&$-\frac{4183}{4860}$&$\frac{8}{9}$&$\frac{2}{9}$\\ [2pt]
				&$1^{++}:{\epsilon^{\mu\alpha\rho\sigma} \epsilon_\rho p_\sigma p^\nu \atop -\{\mu\leftrightarrow\nu\}}$&$\frac{11}{48}$&$\frac{5}{16}$&$-\frac{1}{27}$&$-\frac{8}{9}$&$0$&$\frac{7}{54}$&$\frac{25}{108}$&$-\frac{1}{3}$&$-\frac{8141}{38880}$&$\frac{4}{9}$&$\frac{1}{18}$\\[2pt]
				&$2^{++}:\epsilon^{\mu\alpha}p^\nu - \epsilon^{\nu\alpha}p^\mu$&$\frac{13}{24}$&$\frac{1}{24}$&$-\frac{2}{3}$&$-\frac{16}{9}$&$\frac{1}{24}$&$\frac{8}{27}$&$\frac{10}{27}$&$\frac{7}{12}$&$-\frac{8483}{19440}$&$-\frac{8}{9}$&$-\frac{1}{9}$\\
				&$2^{-+}:\epsilon^{\mu\nu\rho\sigma}{\epsilon_\rho}^\alpha p_\sigma$&$-\frac{13}{24}$&$\frac{1}{24}$&$-\frac{2}{3}$&$\frac{16}{9}$&$\frac{\
					1}{24}$&$-\frac{8}{27}$&$\frac{10}{27}$&$-\frac{7}{12}$&$-\frac{8483}{\
					19440}$&$-\frac{2}{9}$&$-\frac{1}{9}$
			\end{tabular}
		\end{ruledtabular}
	\end{table*}
\end{turnpage}


\break

\bibliography{refs}

\end{document}